\documentstyle[12pt]{article}\textheight
220mm\textwidth 160mm 
\newcommand{\bi}{\bibitem}
\newcommand{\be}{\begin{eqnarray}}
\newcommand{\ee}{\end{eqnarray}}
\newcommand{\nn}{\nonumber}
\catcode`\@=11
\def\lsim{\mathrel{\mathpalette\@versim<}}
\def\gsim{\mathrel{\mathpalette\@versim>}}
\def\@versim#1#2{\vcenter{\offinterlineskip
\ialign{$\m@th#1\hfil##\hfil$\crcr#2\crcr\sim\crcr } }}
\catcode`\@=12

\begin{document}
\pagestyle{empty}

\noindent
\hspace*{10.7cm} \vspace{-3mm} KEK Preprint 97-103\\
\hspace*{10.7cm} \vspace{-3mm} KANAZAWA-97-12\\

\vspace{0.3cm}
\noindent
\hspace*{10.7cm} July 1997

\vspace{0.3cm}

\begin{center}
{\Large\bf  Constraints on \vspace{-1mm}
Finite Soft \\ Supersymmetry--Breaking Terms}
\end{center} 

\vspace{1cm}

\begin{center}
{\sc Tatsuo Kobayashi}$\ ^{(1)}$, {\sc Jisuke Kubo}$\ ^{(2),*}$,\vspace{-1mm}\\
 {\sc Myriam
Mondrag\' on}$\ ^{(3),**}$ \vspace{-1mm} and 
{\sc George Zoupanos}$\ ^{(4),***}$  
\end{center}
\begin{center}
{\em $\ ^{(1)}$ Institute of Particle and\vspace{-2mm}
 Nuclear Studies\\
High Energy Accelerator Research  Organization,
 Tanashi,  \vspace{-2mm} Tokyo 188, Japan} \\
{\em $\ ^{(2)}$ 
Department of Physics, 
Kanazawa \vspace{-2mm} University,
Kanazawa 920-1192, Japan } \\
{\em $\ ^{(3)}$ Instituto de F\' isica,  \vspace{-2mm} UNAM,
Apdo. Postal 20-364,
M\' exico 01000 D.F., M\' exico}\\
{\em $\ ^{(4)}$ 
Instit\" ut f\" ur Physik, Humboldt-Universit\" at zu Berlin, \vspace{-2mm}
D-10115 Berlin, Germany}
\end{center}

\begin{center}
{\sc\large Abstract}
\end{center}

\noindent
Requiring the soft supersymmetry-breaking (SSB) parameters 
in finite \vspace{-2mm} gauge-Yukawa unified models to be
finite up to  and including two-loop order, \vspace{-2mm} we derive
a two-loop sum rule for the
soft scalar-masses.
It is shown that this sum rule \vspace{-2mm} coincides
with that of a certain class of 
string models  in which the massive string states are \vspace{-2mm}
organized into $N=4$ supermultiplets.
We  investigate \vspace{-2mm}
the SSB sector  of two finite $SU(5)$ models.
Using the sum rule which allows
the non-universality of the SSB
terms and requiring that \vspace{-2mm}
the lightest
superparticle particleis  neutral, we 
constrain the  parameter space \vspace{-2mm}
of the
SSB sector in each model.

\vspace*{1cm}
\footnoterule
\vspace*{2mm}
\noindent
$^{*}$Partially supported  by the Grants-in-Aid
for \vspace{-3mm} Scientific Research  from the Ministry of
Education, Science 
and Culture \vspace{-3mm}  (No. 40211213).\\
\noindent
$ ^{**}$
Partially supported \vspace{-3mm} by the UNAM Papiit project IN110296. \\
\noindent
$ ^{***}$
On leave from:
Physics Dept., Nat. Technical University, GR-157 80 
\vspace{-3mm} Zografou,\\
Athens, Greece.
Partially supported  by the E.C. projects, \vspace{-3mm}
FMBI-CT96-1212 and  ERBFMRXCT960090,
the Greek projects, PENED95/1170; 1981.

\newpage
\pagestyle{plain}
\section{Introduction}
The standard model (SM)  has a large number
of free parameters whose values are determined only experimentally. 
To reduce the number of these free parameters, and thus render
it more predictive, one is usually led to enlarge the symmetry
of the SM. 
For instance,  unification of the SM forces based on the  $SU(5)$ GUT 
\cite {gut1} was predicting
one of the  gauge couplings \cite{gut1} as well as the mass of 
the bottom quark \cite{begn}.
Now it seems that LEP data is suggesting that the symmetry of the
unified theory should be further enlarged and become  $N=1$ globally
supersymmetric  \cite{abf}.

Relations among gauge and Yukawa couplings,
which are missing in ordinary GUTs,
could be a consequence of a further unification provided by a more 
fundamental theory
at the Planck scale. Moreover,  it might be possible that
some of these relations are renormalization group invariant (RGI)
below the  Planck scale so that they  
are exactly preserved down to the GUT scale $M_{\rm GUT}$.
In fact, one of
our motivation in this paper is
to point out such indication
in the soft supersymmetry-breaking (SSB) sector in supersymmetric
unified theories.

In our recent studies \cite{kmz1}-\cite{kmoz2},
we have been searching for
RGI relations among gauge and Yukawa couplings in various
supersymmetric GUTs.
Thus, the  idea of gauge-Yukawa unification (GYU)
\cite{kmz1}-\cite{kmoz2} relies  not only  
on a symmetry principle, but also on the principle of
reduction of couplings \cite{zim1,kubo1}
(see also \cite{chang1}). This  principle is based on the
existence of RGI relations among couplings, which 
do not necessarily result from a symmetry,
but nevertheless preserve perturbative renormalizability
or even finiteness.
Here we would like to focus on finite
unified theories  
\cite{parkes1}-\cite{yoshioka1}, \cite{kmz1}, \cite{kmoz2}. 

Supersymmetry seems  to be essential for a successful GYU,
but, as it is for any realistic  supersymmetric model,
the breaking of supersymmetry
has to be understood. We recall that the SSB
parameters have dimensions greater than or equal to one and it is
possible to treat
dimensional couplings
along the line of GYU \cite{piguet1,jack1}, which shows that
the SSB sector of a GYU model
is  controlled by the unified gaugino mass $M$.
As for one- and two-loop finite SSB terms, only
the universal solution for the SSB terms 
\cite{parkes1,jack3} is known so far.
So another motivation of this paper is 
to re-investigate the conditions
for the two-loop finite SSB terms and to
express  them in terms of 
simple sum rules for these parameters.
We will indeed find that the universal
solution can be relaxed for  the SSB terms
to be finite  up to and
including the two-loop corrections, and 
we will derive the two-loop corrected 
sum rule for the soft scalar-masses.
We will comment on the possibility
of all-order-finite SSB terms.

The authors of \cite{ibanez1,munoz1,jack1} have pointed out that
the universal soft scalar masses 
also appear for dilaton-dominated supersymmetry breaking in 4D superstring
models \cite{IL}-\cite{BIM}.
Ib\' a\~ nez \cite{ibanez1} 
(see also \cite{munoz1}) gives a possible superstring interpretation
to it.
We shall examine whether or not the 
two-loop corrected sum rule 
can also be obtained in some string model. We will indeed
find that there is a class of 4D orbifold models
in which  exactly the same sum rule is satisfied.
It may be worth-mentioning that not only in finite GYU models,
but also in nonfinte GYU models  the same soft scalar-mass
sum rule is satisfied at the one-loop level \cite{kkk}.
In ref. \cite{kkk} a possible answer to why this happens is speculated.

Motivated by the fact that
 the universal choice for the
SSB terms can be relaxed,
we will investigate
the SSB sector of two finite $SU(5)$ models.
The SSB parameters of these models are constrained by
the sum rule and also by the 
requirement that the electroweak gauge symmetry
is radiatively broken \cite{inoue1}.
We will find that there is a parameter range
for each model in which the lightest superparticle (LSP)
is a neutralino, which will be compared 
with the case of the universal SSB parameters.
The lightest Higgs turns out to be
$\sim 120$ GeV.

\newpage

\section{Two-loop finiteness and Soft scalar-mass sum rule}
\subsection{Two-loop finite SSB terms}
Various groups \cite{rgf,jack3} have independently computed
the coefficients of the two-loop RG functions
for SSB parameters \footnote{The RG 
functions \cite{jones2,PW,rgf,jack1,jack3} are given in Appendix for
completeness.}. Here we would like to use them to re-investigate
their two-loop finiteness and derive
the two-loop soft scalar-mass sum rule.

The superpotential is
\be
W &=&\frac{1}{6} \,Y^{ijk}\,\Phi_i \Phi_j \Phi_k
+\frac{1}{2} \,\mu^{ij}\,\Phi_i \Phi_j~,
\ee
along with the Lagrangian for SSB terms,
\be
-{\cal L}_{\rm SB} &=&
\frac{1}{6} \,h^{ijk}\,\phi_i \phi_j \phi_k
+
\frac{1}{2} \,b^{ij}\,\phi_i \phi_j
+
\frac{1}{2} \,(m^2)^{j}_{i}\,\phi^{*\,i} \phi_j+
\frac{1}{2} \,M\,\lambda \lambda+\mbox{H.c.}~
\ee
Since we would like to consider
only finite theories here, we assume that 
the gauge group is  a simple group and the one-loop
$\beta$ function of the 
gauge coupling $g$ (A.1) vanishes, i.e.,
\be
b \equiv T(R)-3 C(G)=0~.
\label{Q}
\ee
We also assume that the reduction equation
\be
\beta_{Y}^{ijk} &=& \beta_{g}\,d Y^{ijk}/d g
\label{Yg2}
\ee
admits power series solutions of the form
\be 
Y^{ijk} &=& g\,\sum_{n=0}\,\rho^{ijk}_{(n)} g^{2n}~,
\label{Yg}
\ee
where $\beta_{g}$ and $ \beta_{Y}^{ijk}$
are $\beta$ functions of $g$ and $Y^{ijk}$, respectively.
According to the finiteness theorem of ref. \cite{LPS},
the theory is then finite \footnote{Finiteness here
means only for dimensionless couplings, i.e.
$g$ and $Y^{ijk}$.}
to all orders in perturbation theory,
if the one-loop anomalous dimensions $\gamma_{i}^{(1)\,j}$ 
given in (A.2) vanish, i.e., if
\be
 \frac{1}{2}\sum_{p,q}\rho_{ipq(0)} \rho^{jpq}_{(0)}
-2\delta^{j}_{i}\,C(i) &=&0~
\label{P}
\ee
is satisfied, where we have inserted $Y^{ijk}$ in (\ref{Yg2})
into $\gamma_{i}^{(1)\,j}$.
We recall that if the conditions (\ref{Q}) and (\ref{P})
are satisfied, the two-loop expansion coefficients
in (\ref{Yg}), $\rho^{ijk}_{(1)}$, vanish \cite{jack3}. (From 
(A.6) ad (A.7)
we see that the two-loop coefficients $\beta_{g}^{(2)}$ and
$\gamma_{j}^{(2)i}$ vanish if  $\beta_{g}^{(1)}$ and $\gamma_{j}^{(1)i}$
vanish.)
Further, the one- and two-loop finiteness for $h^{ijk}$ 
can be achieved by \cite{jones2,jack3}
\be
h^{ijk} &=& -M Y^{ijk}+\dots =-M \rho^{ijk}_{(0)}\,g+O(g^5)~,
\label{hY}
\ee
which can be seen from (A.9) if one uses eq. (\ref{P}).
Note further that the $O(g^3)$ term is absent in (\ref{hY}).
As for $b^{ij}$ there is no constraint;
$b^{ij}$ is finite if eqs. (\ref{P}) and (\ref{hY})
are satisfied, which can be seen from the
one- and two-loop coefficients of the $\beta$
function for $b^{ij}$(A.5) and (A.10).

Now, to obtain the two-loop sum rule for 
soft scalar masses, we assume that 
the lowest order coefficients $\rho^{ijk}_{(0)}$ 
and also $(m^2)^{i}_{j}$ satisfy the diagonality relations
\be
\rho_{ipq(0)}\rho^{jpq}_{(0)} &\propto & \delta_{i}^{j}~\mbox{for all} 
~p ~\mbox{and}~q~~\mbox{and}~~
(m^2)^{i}_{j}= m^{2}_{j}\delta^{i}_{j}~,
\label{cond1}
\ee
respectively.
Then one finds that
\be
[\beta_{m^2}^{(1)}]^{j}_{i} &=& \rho_{ipq(0)}
\rho^{jpq}_{(0)}\,(~m_{i}^{2}/2+m_{j}^{2}/2
+m_{p}^{2}+m_{q}^{2}~)g^2 \nn\\
& +& (\rho_{ipq(0)}\rho^{jpq}_{(0)}-8\delta^{j}_{i} C(i))\,M M^{\dag}g^2   
+O(g^6)
\label{betam}~,
\ee
where we have used $\rho^{jpq}_{(1)}=0$
(which implies that the $O(g^4)$ term in (\ref{betam}) is absent).
Using the condition
(\ref{P}), the diagonality relations (\ref{cond1}) and also
the soft scalar-mass sum rule (which we are going to prove)
\be
(~m_{i}^{2}+m_{j}^{2}+m_{k}^{2}~)/
M M^{\dag} &=&
1+\frac{g^2}{16 \pi^2}\,\Delta^{(1)}+O(g^4)~
~\mbox{for}~~i,j,k~~\mbox{with}~\rho^{ijk}_{(0)} \neq
0, \label{sumr} 
\ee
we  find that eq. (\ref{betam}) can be written as
\be
[\beta_{m^2}^{(1)}]^{j}_{i} &=&
4\delta^{j}_{i} \,M M^{\dag} C(i) \,\Delta^{(1)}
\frac{g^4}{16 \pi^2}+O(g^6)~.
\label{betam1}
\ee
We will find shortly that the two-loop correction term $\Delta^{(1)}$
is  given by
\be
\Delta^{(1)} &=&  -2\sum_{l} [(m^{2}_{l}/M M^{\dag})-(1/3)]~T(R_l)~.
\label{delta}
\ee
Therefore, the $\Delta^{(1)}$ vanishes
for the universal choice
\be
m_{i}^{2} &=&\kappa_{i}M M^{\dag}~~
\mbox{with}~~\kappa_{i}=\frac{1}{3}~~\mbox{for all}~~i~,
\label{sym}
\ee
in accord with the previous findings of refs. \cite{jack3}.
The result agrees also with that of ref. \cite{parkes1} on $N=4$
theory;  $N=4$
theory contains three  $N=1$ chiral superfields in the adjoint
representation, which means $T(R_i)=C(G)~~(i=1,2,3) $. If
$\kappa_{1}+
\kappa_{2}+\kappa_{1}=1$ is satisfied, we obtain 
\be
\Delta^{(1)}(N=4) & =&  -2\sum_{l=1}^3 
[\kappa_l-(1/3)]~C(G)=0~.
\ee

To see that $\Delta^{(1)}$ is really given by
eq. (\ref{delta}) for two-loop finiteness
of $m_{i}^{2}$, we recall that 
the two-loop $\beta$ function 
for $m_{i}^{2}$  (A.11) can be
nicely organized as \cite{jack1}
\be
[\beta_{m^2}^{(2)}]^{j}_{i} &=&
(~ A^{jp}_{(\gamma)in}~\gamma^{(1)n}_{p}+
A^{jp}_{(m^2)in}~[\beta_{m^2}^{(1)}]^{n}_{p}+
A_{(g)i}^{j} ~\beta_{g}^{(1)}\nn\\
& &+A^{jp}_{(h)in}~[\, h^{nrq} Y_{lrq}
+4M\delta^{n}_{p}\, g^2\,C(n)\,]
+4 g^4 C(i) S' M M^{\dag}\delta_{i}^{j} ~ ) +\mbox{H.c.},
\label{betam2}
\ee
where 
\be
S' &=& \sum_{l} (m^{2}_{l}/M M^{\dag})~T(R_l) -C(G)~\nn\\
&=& \sum_{l} [(m^{2}_{l}/M M^{\dag})-(1/3)]~T(R_l)~
~\mbox{for}~~\sum_{l} T(R_l)=3 C(G)~,
\label{S}
\ee
and  
 the coefficients $A$'s  are given in (A.11).
Using the one-loop finiteness conditions 
(which are ensured by eqs.
(\ref{Q}), (\ref{P}), (\ref{hY}) and (\ref{sumr})), we finally  obtain
\be
[\beta_{m^2}^{(2)}]^{j}_{i} &=&
+8 g^4 C(i) M M^{\dag}
S'~\delta_{i}^{j}~.
\ee
It is  now easy to see that this term can be canceled
by the $O(g^4)$ contribution to $[\beta_{m^2}^{(1)}]^{j}_{i}$
(which is given in (\ref{betam1})) if
$\Delta^{(1)}$ is exactly given by eq. (\ref{delta}).
Note that we have not shown that the sum rule (\ref{sumr})
is the unique solution for 
$[\beta_{m^2}^{(2)}]^{j}_{i}$.
That is, we have only shown that the sum rule (\ref{sumr})
is a solution to
\be
\rho_{ipq(0)}
\rho^{jpq}_{(0)}\,~[~(~m_{i}^{2}
+m_{p}^{2}+m_{q}^{2}~)/|M|^2-1~]
&=& -8 S' \delta_{i}^{j}C(i)~,
\label{unique}
\ee
but not in the opposite way.
The question  of whether the
sum rule is the unique solution to (\ref{unique}) depends
on the concrete model of course.
We will address the question when discussing concrete
finite models and  find that
the sum rule (\ref{sumr}) is the unique solution
for these models.

Since  $S'$ will be of $O(C(G))$, the two-loop correction
term in the sum rule (\ref{sumr}) may be estimated as
\be
\frac{g^2}{16\pi^2} \Delta^{(1)} 
\sim \frac{\alpha_{\rm GUT}}{\pi}\,C(G)~.
\ee
If, however, the soft scalar masses are close to the universal
one (\ref{sym}), the correction is small.
In the concrete  example of the $SU(5)$
finite models which we will consider below, it will turn out that
the soft scalar masses should differ from
the universal one 
if we require that  the LSP is a neutralino.
But the two-loop correction term $\Delta^{(1)}$
happens to vanish exactly, no matter how large
the deviation from the universal choice
of the soft scalar masses is.

\subsection{Coincidence}
It has been known \cite{jack1,ibanez1,munoz1} that
the universal soft scalar masses 
which preserve their  two-loop finiteness 
also appear for dilaton-dominated 
supersymmetry breaking in 4D superstring
models \cite{IL}-\cite{BIM}.
Ib\' a\~ nez \cite{ibanez1} 
(see also \cite{munoz1}) gives a possible superstring interpretation
and argues that for dilaton dominance to work,
the soft SSB terms have to be 
 be independent of the particular
choice of compactification and 
consistent with any
possible compactification, in  particular
with a toroidal compactification
preserving $N=4$ supersymmetry.
Given that 
 the universality of the soft scalar masses 
can be relaxed 
(as we have shown above),
we would like to examine whether or not the 
two-loop corrected sum rule (\ref{sumr})
can also be obtained in some string model.
To this end, we
 consider a specific class of
orbifold models with three untwisted moduli $T_1,T_2,T_3$
(which exist  for instance in $(0,2)$ symmetric abelian
orbifold construction always).
We then 
 assume that 
 some non-perturbative superpotential which breaks supersymmetry
exists and that the dilation $S$ and the moduli $T_a$ play
the dominant role for supersymmetry breaking.
The K\" ahler potential $K$  and 
the gauge kinetic function $f$ in this case 
assume the  generic form
\begin{eqnarray}
&~& K=-\ln (S+S^*) -\sum_{a=1}^{3}\, \ln
(T_a+T^{*}_{a})+\sum_i\,
\Pi_{a=1}^{3}(T_a+T^{*}_{a})^{n_{i}^{a}}|\Phi_i|^2,~f=k S,
\label{K-string} 
\end{eqnarray}
where $n_{i}^{a}$ stand for   modular weights and are fractional 
numbers, and  $k$ is
the Kac-Moody level  \cite{witten2}-\cite{ferrara3}.
The SSB parameters \footnote{Since the SSB parameter
$b^{ij}$are not constrained by  two-loop finiteness,
we do not consider it here.}
in this class of models are given by
\cite{BIM}, \cite{multiT}-\cite{SD}, \cite{ibanez1,munoz1}
\be
M &=& \sqrt 3 m_{3/2} \sin \theta~,~
m_i^2 = m_{3/2}^2(1+3\cos^2 \theta \,
\sum_{a=1}^{3} n_i^a \Theta^2_a ) \label{stringm}~,\\
h^{ijk} &=& -\sqrt 3 Y^{ijk}m_{3/2}[\sin \theta 
+\cos \theta \sum_{a=1}^{3} \Theta_a (u^a+n_i^a+n_j^a+n_k^a))]~,
\label{stringm1} 
\ee
where $\theta$ and $\Theta_a$ 
(which parametrize the unknown
 mechanism of supersymmetry breaking \cite{BIM}) are defined
as $ F^S /Y =\sqrt{3} m_{3/2} \sin\theta$
and \newline $ F^{T_a}/(T_a+
T_{a}^{*})=\sqrt{3} m_{3/2} \cos\theta
\Theta_a$ with $\sum_{a=1}^{3} \Theta_{a}^{2}=1$.
In eq.~(\ref{stringm1}) we have assumed $Y^{ijk}$ is independent of $S$ 
and $T_a$.
It is straightforward 
 to see that 
the tree-level for of the sum rule
(\ref{sumr}) \cite{BIM,BIM2,ibanez1,munoz1,kim1}\footnote{We call the 
soft scalar-mass sum rule
(\ref{sumr}) without the
two-loop correction term the tree-level sum rule.} is
satisfied, if
\be
{\bf n}_i+{\bf n}_j+{\bf n}_k &=& -{\bf u} \equiv -(1,1,1)~.
\label{3n}
\ee
Note that the condition (\ref{3n}) ensures that 
$K+\ln |W|^2$ is invariant under the duality transformation,
\be
T_a \rightarrow {a_aT_a-ib_a \over ic_aT_a + d_a},
\ee
where $a_a$, $b_a$, $c_a$ satisfying
 $d_a$ are integers and $a_ad_a-b_ac_a=1$.
The  K\" ahler potential $K$ (\ref{K-string})
belongs to the general class of the K\"ahler
potentials that lead to the tree-level 
sum rule \cite{kkk}.
When gauge symmetries break, we generally have $D$-term contributions 
to the soft scalar masses.
Such $D$-term contributions, however, do not appear in the sum rule,
because each $D$-term contribution is proportional to the charge of 
the matter field $\Phi_i$ \cite{Dterm}.

We then would like to extend our discussion
so as to include the two-loop correction in the
sum rule (\ref{sumr}).
In superstrings, the correction to the tree level relations
among the SSB terms 
can be computed by using the fact
that the target-space modular anomaly 
\cite{ferrara1,lopes1,IL} 
 are canceled by the Green-Schwarz mechanism \cite{GS} and
the threshold correction coming from the massive 
sates \cite{kap2,DKL2}.
The Green-Schwarz mechanism
induces a nontrivial transformation of $S$ under
th duality transformation, which implies that the K\"ahler
potential for the dilaton $S$  
has to be modified to
the duality-invariant K\"ahler
potential \cite{ferrara1,IL}, 
\begin{equation}
- \ln  Y~,~Y \equiv S+S^*-
\sum_{a=1}^{3}{\delta_{\rm GS}^{a} \over 8\pi^2} 
\ln (T_a+T^{*}_{a})~,
\label{Kahler}
\end{equation}
where $\delta_{\rm GS}^{a}$ is 
the Green-Schwarz coefficient \cite{ferrara1,IL}.
This correction alters the tree-level formulae for
$h^{ijk}$ and $m_{i}^{2}$, while
the threshold correction coming from the massive sates
modifies the tree-level gauge kinetic function $f=S$
and hence changes the tree-level formula for the gaugino mass $M$.
The requirement of the vanishing cosmological constant 
leads to
the redefinition of the Goldstino 
parameters \cite{multiT}-\cite{SD} as
\be
\frac{1}{Y}(\,F^S-\sum_a \, \frac{\delta_{\rm GS}^a /8 \pi^2}
{T_a+T^{*}_{a}}\,) F^{T_a} &=&\sqrt{3} m_{3/2} \sin\theta ~,\\
\frac{F^{T_a}}{T_a+T^{*}_{a}}\,
&=& \sqrt{3} m_{3/2} \cos\theta \tilde{\Theta}_a~,
\label{YY}
\ee
where 
\be
\tilde{\Theta}_a &=&
(1-\delta_{\rm GS}^a
 /Y 8 \pi^2)^{-1/2}\,\Theta_a~,
\ee
and $\Theta_a$ is defined in (\ref{stringm1}).
Note that the quantum modification (\ref{YY}) does not change
the tree-level relation
for $h^{ijk}$ (\ref{stringm1}) at all,
which coincides the two-loop result (\ref{hY}).
This motivates us to assume that 
the relation for $M$ also remains unchanged,
which is true only if the contribution
to
the gauge kinetic function $f$ coming from 
the massive states \cite{DKL2} are absent.
It is known \cite{DKL2} (see also \cite{IL})
that such situation 
 appears 
for  the class of orbifold
models in which the massive states are organized into $N=4$
supermultiplets 
\footnote{The absence of the
threshold effects coming from  $N=4$
massive supermultiplets has been first observed in
an $N=4$ Yang-Mills theory with spontaneously
broken gauge symmetry \cite{kuboM}.} and we obtain
One can easily convince oneself that if
the condition (\ref{3n}) is satisfied, 
the tree-level sum rule for $m_{i}^{2} $ is 
modified to 
\be
\frac{m_{i}^{2}+m_{j}^{2}+m_{k}^{2}}{|M|^2}-1
&=& \frac{\cos^2\theta}{\sin^2\theta}\,(1-\sum_{a=1}^{3}
\tilde{\Theta}^{2}_{a})~.
\label{sumr1}
\ee
In this case the duality anomaly should be 
canceled only by the Green-Schwarz mechanism, 
implying that \cite{ferrara1,IL}
\be
\delta_{\rm GS}^{a} &=&
-C(G)+\sum_l\,T(R_l) (1+2 n^{a}_{l})~.
\ee
After a straightforward calculation one then finds 
two identities:
\be
(1-\sum_{a=1}^{3} 
\tilde{\Theta}_{a}^{2} )[1-\frac{g^2}{24\pi^2}\sum_l T(R_l)] &=&
-\frac{g^2}{16\pi^2} [~
2\sum_l\,T(R_l)(\frac{1}{3}+\sum_{a=1}^{3} 
\tilde{\Theta}_{a}^{2} n^{a}_{l})~\nn\\ 
& & +\frac{b}{3}\sum_{a=1}^{3}
\tilde{\Theta}_{a}^{2} 
~]~,\\
\sum_l\,T(R_l)(\frac{m_{l}^{2}}{|M|^2}-\frac{1}{3})
&=& \frac{\cos^2 \theta}{\sin^2\theta}
~\sum_l\,T(R_l)(\frac{1}{3}+\sum_{a=1}^{3} 
\tilde{\Theta}_{a}^{2} n^{a}_{l})~,
\ee
where we have used $Y=2/g^2$. 
Using these identities, one can convince oneself that
the two-loop corrected sum rule (\ref{sumr})
coincides with  the sum rule (\ref{sumr1})  of the orbifold
model up to and including $O(g^2)$ terms.
For finite theories $(b=0)$ it is possible
to  express the sum rule (\ref{sumr1}) in terms of field
theory quantities only:
\be
\frac{m_{i}^{2}+m_{j}^{2}+m_{k}^{2}}{|M|^2}-1
&=& \frac{\sum_l\,T(R_l)(m_{l}^{2}/|M|^2-1/3)}
{C(G)-8 \pi^2/g^2}~.
\label{sumr2}
\ee
It is remarkable that in this combination of the SSB
terms the quantities such as the
Goldstino angle parameterizing  unknown supersymmerty breaking
disappear.
Since the sum rule (\ref{sumr2}) can be seen
as an exact result,
we conjecture that the sum rule (\ref{sumr2}) 
and the tree-level form of the relation  $h^{ijk}=
-M Y^{ijk}(g)$ are also exact 
 results in field theory that
result from the finiteness of the SSB
parameters. 

\subsection{Comment}
We next would like to comment on the possibility of
having all-order finite SSB terms.
To begin with we recall that the RG functions are 
renormalization-scheme dependent starting at two-loop order.
This is true, even if we assume that a mass-independent
 renormalization-scheme is employed, except for the gauge coupling
$\beta$ function.
Therefore,  it could be possible to find a renormalization-scheme
in which all the higher order coefficients of the $\beta$ functions
 (except for the gauge coupling $\beta$ function) vanish.
Since we know explicitly the two-loop RG functions,
we would like to ask whether we can find a renormalization scheme
in which all the RG functions beyond the two-loop vanish.
To simplify the problem, we assume that
all the supersymmetric, massive parameters are set equal to zero
and that $Y^{ijk}$ and $h^{jik}$ have been reduced in favor of
$g$ and $M$. 
Suppose that we have found reparametrizations of 
$g$ , $M$ and $m^{2}$
such that the $\beta$ functions, except for $\beta_{g}$ and
$\beta_{m^2}$,
beyond the two-loop order vanish.
We then ask ourselves whether or not it 
is possible to find a reparametrization of $m^{2}_{i}$'s  of the form
\be
m^{2}_{i} & \to & m^{2}_{i}+\frac{g^4}{16\pi^2} K_i~~
\mbox{with}~~
K_i=r_{ij} m_{j}^{2} +p_i |M|^2~, 
\label{rp}
\ee
where $ r_{ij}$ and $p_i$ are numbers, 
such that the three-loop $\beta$ functions 
for $m^{2}_{i}$'s vanish \footnote{It is possible to
find a reparametrization of $m^{2}_{i}$
and then to make $\beta_{m^2}^{(2)}$ zero.}.
Inserting (\ref{rp}) into the one-loop $\beta$ function (A.4), 
we see that the
three-loop terms in the $\beta$ function
 should be canceled by the term
\be
\rho_{ipq(0)} \rho^{jpq}_{(0)} (K_i+
K_j+K_k) ~,
\label{Kj}
\ee
where we have used eq. (\ref{Yg2}).
Recall that because of the diagonality condition (\ref{cond1})
the terms given above are proportional to $\delta_{i}^{j}$
and so the total number of these terms, $N$, is exactly
the number of the chiral superfields present in the theory.
It is clear that if 
 these $N$ terms are linearly independent,
the three-loop contributions in the $\beta$ functions for 
$m^{2}_{i}$'s can be  canceled by them.

This algebraic question is very much related to
the question of whether or not the sum rule
is the unique solution to the two-loop finiteness,
because it depends on the explicit form of $\rho^{ijk}_{(0)}$.
One can convince oneself that
if the sum rule
is the unique solution to the two-loop finiteness
and the sum rule does not fix  $m^{2}_{i}/|M|^2$
completely, the $N$ terms given in (\ref{Kj}) 
are not linearly independent. In
this case, it is not clear from the beginning that three-loop terms in the
$\beta$ function can be canceled by (\ref{Kj}); 
one has to compute explicitly
the three-loop contributions to see it.
In the concrete models we will consider later, these $N$ terms (\ref{Kj})
are not linearly independent.
The string inspired result (\ref{sumr2}) should have
a nontrivial meaning in this case; it suggests
that the three-loop contributions
can be canceled by a reparametrization of $m^{2}_{i}$, because
the reparametrization defined by
\be
m^{2}_{i} &\to & m^{2 \prime}_{i} = m^{2}_{i}-
\frac{1}{3}
\frac{\sum_l\,T(R_l)(m_{l}^{2}-|M|^2 /3)}
{C(G)-8 \pi^2/g^2}
\ee
can bring the ''exact'' result (\ref{sumr2})
into the tree-level
form.
If, on the other hand, 
 the sum rule
is the unique solution to the two-loop finiteness
and the sum rule  fixes  $m^{2}_{i}/|M|^2$
completely, the $N$ terms (\ref{Kj}) are linearly independent.
We can then cancel all the three-loop contributions,
which then can be continued to arbitrary order.

\section{Finite theories based on $SU(5)$}
\subsection{General comments}
 
{}From the classification of theories with vanishing one-loop 
gauge $\beta$
function \cite{HPS1}, one can easily see that there exist only two candidate
possibilities to construct $SU(5)$ GUTs with three generations. These
possibilities require that the theory should contain as matter fields the
chiral supermultiplets ${\bf 5}~,~\overline{\bf 5}~,~{\bf 10}~,
~\overline{\bf 5}~,~{\bf 24}$
with the multiplicities $(6,9,4,1,0)$ and
 $(4,7,3,0,1)$, respectively. Only the second one contains a
${\bf 24}$-plet which can be used to provide the spontaneous
 symmetry breaking
(SB) of 
SU(5) down to $SU(3)\times SU(2)\times U(1)$. For the first model one has to
incorporate another way, such as the Wilson flux breaking mechanism to
achieve the desired SB of $SU(5)$ \cite{kmz1}. Therefore, for a
self-consistent field theory discussion we would like to concentrate only on
the second possibility.

 It is clear, at least for the dimensionless couplings, that the matter
content of a theory is only a necessary condition for all-order
finiteness. Therefore, there exist, in principle, various finite models for
a given matter content. However, during the early studies
 \cite{HPS2,mode2},
the theorem \cite{LPS} that guarantees all-order finiteness 
and requires the
existence of power series solution to any finite order in perturbation
theory was not known. The theorem introduces new constraints, in particular
requires that the solution to the one-loop
finiteness conditions should be non-degenerate and 
isolated. In most studies the freedom resulted as a consequence
of the degeneracy in the one- and two-loop solutions has been used to make
specific ans\" atze that could lead to phenomenologically acceptable
predictions.  Note that the existence of such freedom is incompatible with
the power series solutions \cite{zim1,LPS}.

 Taking into account the new constraints an all-order finite $SU(5)$ model
has been constructed \cite{kmz1}, which among others successfully predicted
the bottom and the top quark masses \cite{kmz1,kmoz2}.
The later is due to the
Gauge-and-Yukawa-of-the-third-generation Unification 
\cite{kmz1}-\cite{kmoz2} which has been achieved.
 In general the predictive power of a finite $SU(5)$ model depends on the
structure of the superpotential and on the way the four pairs of Higgs
quintets and 
anti-quintets mix to produce the two Higgs doublets of the
minimal supersymmetric standard model (MSSM). 
Given that the finiteness conditions do not restrict the mass terms,
there is a lot of freedom offered by this sector of the theory in mixing
the four pairs of Higgs fields. As a result it was possible in the early
studies (a) to provide the adequate doublet-triplet splitting in the pair
of ${\bf 5}$ and  $\overline{{\bf 5}}$ 
which couple to ordinary fermions so as to suppress the
proton decay induced by the coloured triplets and (b) to introduce angles
in the gauge-Yukawa relations suppressing in this way the strength of the
Yukawa couplings. Concerning the requirement (b) one has to recall that at
that time it was very unpleasant to have a top mass prediction at
$O(150-200)$ GeV; the popular top quark mass was at $O(40)$ GeV.
The above was
most clearly stated in ref. \cite{mode2} and has been 
revived \cite{yoshioka1}
taking into account the recent data. However, it is clear that using the
large freedom offered by the Higgs mass parameter space in requiring the
condition (b) one strongly diminishes the beauty of a finite
theory. Consequently, this freedom was abandoned in the recent studies of
the all-loop finite SU(5) model \cite{kmz1}
 and only the condition (a) was kept
as a necessary requirement.

\subsection{Models}
 A predictive Gauge-Yukawa unified SU(5) model which is
finite to all orders, in addition to the requirements mentioned already, 
 should also have the following properties.

\begin{enumerate}

\item 
One-loop anomalous dimensions are diagonal,
i.e.,  $\gamma_{i}^{(1)\,j} \propto \delta^{j}_{i} $,
according to the assumption (\ref{cond1}).

\item
Three fermion generations, $\overline{\bf 5}_{i}~~
(i=1,2,3)$, obviously should not couple to ${\bf 24}$.
This can be achieved for instance by imposing $B-L$ 
conservation.

\item
The two Higgs doublets of the MSSM should mostly be made out of a 
pair of Higgs quintet and anti-quintet, which couple to the third
generation.
\end{enumerate}
In the following we discuss two versions of the all-order finite model.

\vspace{0.2cm}
\noindent
${\bf A}$:  The model of ref. \cite{kmz1}.
\newline
${\bf B}$:  A slight variation of  the 
 model ${\bf A}$, which  can also be obtained  from 
the class of the models
suggested by Kazakov {\em et al.} \cite{kazakov1} with a modification to
suppress non-diagonal anomalous dimensions. 

\vspace{0.2cm}
\noindent
The quark mixing can be accommodated in these models,
but for simplicity we
neglect the intergenerational mixing and postpone the interesting problem of
predicting the mixings to a future publication.

The  superpotential which describe the two models 
takes the form \cite{kmz1,kazakov1}
\be
W &=& \sum_{i=1}^{3}\,[~\frac{1}{2}g_{i}^{u}
\,{\bf 10}_i{\bf 10}_i H_{i}+
g_{i}^{d}\,{\bf 10}_i \overline{\bf 5}_{i}\,
\overline{H}_{i}~] +
g_{23}^{u}\,{\bf 10}_2{\bf 10}_3 H_{4} \nn\\
 & &+g_{23}^{d}\,{\bf 10}_2 \overline{\bf 5}_{3}\,
\overline{H}_{4}+
g_{32}^{d}\,{\bf 10}_3 \overline{\bf 5}_{2}\,
\overline{H}_{4}+
\sum_{a=1}^{4}g_{a}^{f}\,H_{a}\, 
{\bf 24}\,\overline{H}_{a}+
\frac{g^{\lambda}}{3}\,({\bf 24})^3~,
\label{super}
\ee
where 
$H_{a}$ and $\overline{H}_{a}~~(a=1,\dots,4)$
stand for the Higgs quintets and anti-quintets.
Given the superpotential $W$,
we  can  compute now the $\gamma$ functions of the model, from 
which we then
compute the $\beta$ functions.
 We find: 
\be
\gamma^{(1)}_{{\bf 10}_1} &=& \frac{1}{16\pi^2}\,
[\,-\frac{36}{5}\,g^2+3\,(g_{1}^{u})^{2}+
2\,(g_{1}^{d})^{2}\,]~,\nn\\
\gamma^{(1)}_{{\bf 10}_2} &=& \frac{1}{16\pi^2}\,
[\,-\frac{36}{5}\,g^2+3\,(g_{2}^{u})^{2}+
2\,(g_{2}^{d})^{2}
+3\,(g_{23}^{u})^{2}+
2\,(g_{23}^{d})^{2}\,]~,\nn\\
\gamma^{(1)}_{{\bf 10}_3} &=& \frac{1}{16\pi^2}\,
[\,-\frac{36}{5}\,g^2+3\,(g_{3}^{u})^{2}+
2\,(g_{3}^{d})^{2}
+3\,(g_{23}^{u})^{2}+
2\,(g_{32}^{d})^{2}\,]~,\nn\\
\gamma^{(1)}_{\overline{{\bf 5}}_1} &=& \frac{1}{16\pi^2}\,
[\,-\frac{24}{5}\,g^2+
4\,(g_{1}^{d})^{2}\,]~,\nn\\
\gamma^{(1)}_{\overline{{\bf 5}}_2} &=& \frac{1}{16\pi^2}\,
[\,-\frac{24}{5}\,g^2+
4\,(g_{2}^{d})^{2}+4\,(g_{32}^{d})^{2}\,]~,\nn\\
\gamma^{(1)}_{\overline{{\bf 5}}_3} &=& \frac{1}{16\pi^2}\,
[\,-\frac{24}{5}\,g^2+
4\,(g_{3}^{d})^{2}+4\,(g_{23}^{d})^{2}\,]~,\nn\\
\gamma^{(1)}_{H_{i}} &=& \frac{1}{16\pi^2}\,
[\,-\frac{24}{5}\,g^2+
3\,(g_{i}^{u})^{2}+
\frac{24}{5}(g_{i}^{f})^2\,]~~i=1,2,3~,\\
\gamma^{(1)}_{\overline{H}_{i}} &=& \frac{1}{16\pi^2}\,
[\,-\frac{24}{5}\,g^2+
4\,(g_{i}^{d})^{2}+
\frac{24}{5}(g_{i}^{f})^2\,]~~i=1,2,3~,\nn\\
\gamma^{(1)}_{H_{4}} &=& \frac{1}{16\pi^2}\,
[\,-\frac{24}{5}\,g^2+
6\,(g_{23}^{u})^{2}+
\frac{24}{5}(g_{4}^{f})^2\,]~,\nn\\
\gamma^{(1)}_{\overline{H}_{4}} &=& \frac{1}{16\pi^2}\,
[\,-\frac{24}{5}\,g^2+
4\,(g_{23}^{d})^{2}+4\,(g_{32}^{d})^{2}
+\frac{24}{5}(g_{4}^{f})^2\,]~,\nn\\
\gamma^{(1)}_{{\bf 24}} &=& \frac{1}{16\pi^2}\,
[\,-10\,g^2+
\sum_{a=1}^{4}(g_{a}^{f})^{2}
+\frac{21}{5}\,(g^{\lambda})^{2}\,]~.\nn
\label{gamma5}
\ee
The non-degenerate and isolated solutions to
$\gamma^{(1)}_{i}=0$ for the models
$\{ {\bf A}~,~{\bf B} \}$ are:
 \be
(g_{1}^{u})^2 &=&\{\frac{8}{5}~,~\frac{8}{5}) \}g^2~,
~(g_{1}^{d})^2 =\{\frac{6}{5}~,~\frac{6}{5}\}g^2~,~
(g_{2}^{u})^2=(g_{3}^{u})^2=\{\frac{8}{5}~,~\frac{4}{5}\}g^2~,\nn\\
(g_{2}^{d})^2 &=&(g_{3}^{d})^2=\{\frac{6}{5}~,~\frac{3}{5}\}g^2~,~
(g_{23}^{u})^2 =\{0~,~\frac{4}{5}\}g^2~,~
(g_{23}^{d})^2=(g_{32}^{d})^2=\{0~,~\frac{3}{5}\}g^2~,
\label{SOL5}\\
(g^{\lambda})^2 &=&\frac{15}{7}g^2~,~
(g_{2}^{f})^2 =(g_{3}^{f})^2=\{0~,~\frac{1}{2}\}g^2~,~
(g_{1}^{f})^2=0~,~
(g_{4}^{f})^2=\{1~,~0~\}g^2~.\nn
\ee
We have explicitly checked that these solutions (\ref{SOL5})
are also the solutions of the reduction equation (\ref{Yg2})
and that they can be uniquely extended to the corresponding
power series solutions 
(\ref{Yg2}) \footnote{The coefficients in 
(\ref{SOL5}) are slightly different from
those  in models considered  in refs.
\cite{kazakov1}.}.
Consequently, these models are finite to all orders.

After the reduction of couplings (\ref{SOL5})
the symmetry of $W$ (\ref{super}) is enhanced:
For the model ${\bf A}$ one finds that
the superpotential has the 
$Z_7\times Z_3\times Z_2$ discrete symmetry 
\be
\overline{{\bf 5}}_{1} &:& (4,0,1)~,~
\overline{{\bf 5}}_{2}: (1,0,1)~,~
\overline{{\bf 5}}_{3}: (2,0,1)~,\nn\\
{\bf 10}_{1} &:& (1,1,1)~,~{\bf 10}_{2}: (2,2,1)~,~
{\bf 10}_{3}: (4,0,1)~,\nn\\
H_{1} &:& (5,1,0)~,~H_{2}: (3,2,0)~,~
H_{3}: (6,0,0)~,\\
\overline{H}_{1} &:& (-5,-1,0)~,~\overline{H}_{2}: (-3,-2,0)~,~
\overline{H}_{3}: (-6,0,0)~,\nn\\
H_{4} &: & (0,0,0)~,~
\overline{H}_{4} : (0,0,0)~,~{\bf 24}:(0,0,0)~,\nn
\label{Z1}
\ee
while for the model ${\bf B}$
one finds 
$Z_4\times Z_4\times Z_4$  defined as
\be
\overline{{\bf 5}}_{1} &:& (1,0,0)~,~
\overline{{\bf 5}}_{2}: (0,1,0)~,~
\overline{{\bf 5}}_{3}: (0,0,1)~,\nn\\
{\bf 10}_{1} &:& (1,0,0)~,~{\bf 10}_{2}: (0,1,0)~,~
{\bf 10}_{3}: (0,0,1)~,\nn\\
H_{1} &:& (2,0,0)~,~H_{2}: (0,2,0)~,~
H_{3}: (0,0,2)~,\\
\overline{H}_{1} &:& (-2,0,0)~,~\overline{H}_{2}: (0,-2,0)~,~
\overline{H}_{3}: (0,0,-2)~,\nn\\
H_{4} &: & (0,3,3)~,~
\overline{H}_{4} : (0,-3,-3)~,~{\bf 24}:(0,0,0)~,\nn
\label{Z2}
\ee
where the numbers in the parenthesis stand for 
the charges under the discrete symmetries.

The main difference of the models
${\bf A}$ and ${\bf B}$ is
that three pairs of Higgs quintets and anti-quintets couple to 
the ${\bf 24}$ for ${\bf B}$ so that it is not necessary  \cite{kazakov1}
to mix
them with $H_{4}$ and $\overline{H}_{4}$ in order to
achieve the triplet-doublet splitting after SB
of $SU(5)$.
This enhances the predicitivity, because
then the mixing of the three pairs of Higgsess
are strongly constrained 
to fit the phenomenology of the first two generations \cite{kazakov1}.

Before we go to present our analysis on
low-energy predictions of the models, 
we would like to discuss the structure of the sum rule for the soft
scalar masses for each case. According to  (\ref{cond1}),
we recall that they are supposed to be diagonal.  From the one-loop 
finiteness for the soft
scalar masses, we obtain
(there are $\{ 10~,~13 \}$ equations for $15$ unknown $\kappa^{(0)}$'s):
\be
\kappa^{(0)}_{H_{i}} & =& 
1-2\kappa^{(0)}_{{\bf 10}_{i}}~,~
\kappa^{(0)}_{\overline{H}_{i}} =
1-\kappa^{(0)}_{{\bf 10}_{i}}
-\kappa^{(0)}_{\overline{{\bf 5}}_{i}}~~(i=1,2,3)~,
\label{suma} \\
\kappa^{(0)}_{H_{4}} &=&\frac{2}{3}-\kappa^{(0)}_{\overline{H}_{4}}~,
~\kappa^{(0)}_{{\bf 24}}=\frac{1}{3}~~\mbox{for}~~{\bf A}~,\nn
\ee
and
\be
\kappa^{(0)}_{H_{1}} & =& 1-2\kappa^{(0)}_{{\bf 10}_{1}}~,~
\kappa^{(0)}_{H_{2}}=
\kappa^{(0)}_{H_{3}}=
\kappa^{(0)}_{H_{4}}=1-2\kappa^{(0)}_{{\bf 10}_{3}}~,\nn\\
\kappa^{(0)}_{\overline{H}_{1}} &=&
1-\kappa^{(0)}_{{\bf 10}_{1}}
-\kappa^{(0)}_{\overline{{\bf 5}}_{1}}~,~
\kappa^{(0)}_{\overline{H}_{2}}=
\kappa^{(0)}_{\overline{H}_{3}}=
\kappa^{(0)}_{\overline{H}_{4}}=
-\frac{1}{3}+2\kappa^{(0)}_{{\bf 10}_{3}}~,
\label{sumb} \\
\kappa^{(0)}_{\overline{{\bf 5}}_{2}} & =&
\kappa^{(0)}_{\overline{{\bf 5}}_{3}}=
\frac{4}{3}-3\kappa^{(0)}_{{\bf 10}_{3}}~,~
\kappa^{(0)}_{{\bf 10}_{2}}=\kappa^{(0)}_{{\bf 10}_{3}}~,
~\kappa^{(0)}_{{\bf 24}}=\frac{1}{3}~~
\mbox{for}~~{\bf B}~,\nn
\ee
where we have defined
\be
\frac{m^{2}_{i}}{|M|^2} &=& \kappa_{i}^{(0)}+
\frac{g^2}{16 \pi^2} \kappa_{i}^{(1)}+\cdots~,~i={\bf 10}_1,
{\bf 10}_2, \dots, {\bf 24}~.
\ee
We then use the solution (\ref{SOL5})  to calculate
the actual value for $S'$ by using eq. (\ref{S}),
which express the two-loop correction to the sum rule.
Surprisingly, it turns out  for both models that
\be
S' &=&0 ~.
\ee
 That is, the one-loop
sum rule in the present models is not corrected in two-loop order.

Next we would like to address the question of whether the sum rule
(\ref{sumr}) is the unique solution to the two-loop finiteness.
To this end, we recall that
 the two-loop finiteness for the
soft scalar masses follows if eq. (\ref{unique}), i.e.
\be
\rho_{ipq(0)} \rho^{jpq}_{(0)} (\kappa_{i}^{(1)}+
\kappa_{p}^{(1)}+\kappa_{q}^{(1)})
&=& 
-8 C(i) \sum_{l} [\kappa_{p}^{(0)}-(1/3)]~T(R_l)=-8 C(i)S'~,
\ee
is satisfied.
There are $15$ equations for $15$ unknown $\kappa^{(1)}$'s.
We find that the solution is not unique;
it can be parameterized by $\{ 7,4 \}$ parameters for a given $S'$
which is zero for the present models.
For instance, 
\be
\kappa^{(1)}_{H_{i}} &= &-2S'-2 \kappa^{(1)}_{{\bf 10}_{i}}~,~
\kappa^{(1)}_{\overline{H}_{i}}=
-2S'-\kappa^{(1)}_{{\bf 5}_{i}}-     
\kappa^{(1)}_{{\bf 10}_{i}}~(i=1,2,3)~, \label{suma1} \\
\kappa^{(1)}_{H_{4}} &= &-\frac{4S'}{3}
-\kappa^{(1)}_{\overline{H}_{4}}~,~
\kappa^{(1)}_{24}=-\frac{2S'}{3}~\mbox{for}~~{\bf A}~,\nn
\ee
and
\be
\kappa^{(1)}_{H_{1}} &= &-2S'-2 \kappa^{(1)}_{{\bf 10}_{1}}~,~
\kappa^{(1)}_{24}=-\frac{2S'}{3}~,
~\kappa^{(1)}_{{\bf 10}_{2}}=\kappa^{(1)}_{{\bf 10}_{3}}~,\nn\\
\kappa^{(1)}_{H_{2}} &=&
\kappa^{(1)}_{H_{3}}=
\kappa^{(1)}_{H_{4}}=-2S'-2 \kappa^{(1)}_{{\bf 10}_{3}}~,~
\kappa^{(1)}_{\overline{H}_{2}}=
\kappa^{(1)}_{\overline{H}_{3}}=
\kappa^{(1)}_{\overline{H}_{4}}=\frac{2S'}{3}
+2 \kappa^{(1)}_{{\bf 10}_{3}}~,
\label{sumb1} \\
\kappa^{(1)}_{\overline{H}_{1}} &=&  
-2S'-\kappa^{(1)}_{\overline{{\bf 5}}_{1}}- \kappa^{(1)}_{{\bf 10}_{1}}~,~
\kappa^{(1)}_{\overline{{\bf 5}}_{2}}=
\kappa^{(1)}_{\overline{{\bf 5}}_{3}}=   
-\frac{8S'}{3}- 3\kappa^{(1)}_{{\bf 10}_{3}}~~\mbox{for}~~{\bf B}~.\nn
\ee
As one can easily see that $\kappa^{(1)}$'s satisfy
\be
\kappa^{(1)}_{i}+\kappa^{(1)}_{j}
+\kappa^{(1)}_{k} &=& -2 S'=0~,
\ee
which shows that the sum rule (\ref{sumr}) in the 
present models is the unique solution
to two-loop finiteness.

\section{Predictions of Low Energy Parameters}

Since the gauge symmetry is spontaneously broken
below $M_{\rm GUT}$, the finiteness conditions 
do not restrict the renormalization property at low energies, and
all it remains are boundary conditions on the
gauge and Yukawa couplings (\ref{SOL5})
and the $h=-MY$ relation (\ref{hY}) and
the soft scalar-mass sum rule (\ref{sumr}) at $M_{\rm GUT}$.
So we examine the evolution of these parameters according
to their renormalization group equations at two-loop 
for dimensionless parameters and 
 at one-loop 
for dimensional ones with
these boundary conditions.
Below $M_{\rm GUT}$ their evolution is assumed to be
governed by the MSSM. We further assume a unique 
supersymmetry breaking scale
$M_{s}$ so that
below $M_{s}$ the SM is the correct effective theory.

We  recall that
$\tan\beta$ is usually determined in the Higgs sector.
However, it has turned out that
in the case of  GYU models it is convenient 
to define $\tan\beta$ by using
the matching condition at $M_{s}$ \cite{barger},
\be
\alpha_{t}^{\rm SM} 
&=&\alpha_{t}\,\sin^2 \beta~,~
\alpha_{b}^{\rm SM}
~ =~ \alpha_{b}\,\cos^2 \beta~,
~\alpha_{\tau}^{\rm SM}
~=~\alpha_{\tau}\,\cos^2 \beta~,\nn\\
\alpha_{\lambda}&=&
\frac{1}{4}(\frac{3}{5}\alpha_{1}
+\alpha_2)\,\cos^2 2\beta~,
\label{match}
\ee
where $\alpha_{i}^{\rm SM}~(i=t,b,\tau)$ are
the SM Yukawa couplings and $\alpha_{\lambda}$ is 
the Higgs coupling $(\alpha_I=g^{2}_{I}/4\pi^2)$.
With a given set of the input
parameters \cite{pdg}, 
\be
M_{\tau} &=&1.777 ~\mbox{GeV}~,~M_Z=91.188 ~\mbox{GeV}~,
\label{mtau}
\ee
with \cite{pokorski1}
\be
\alpha_{\rm EM}^{-1}(M_{Z})&=&127.9
+\frac{8}{9\pi}\,\log\frac{M_t}{M_Z} ~,\nn\\
\sin^{2} \theta_{\rm W}(M_{Z})&=&0.2319
-3.03\times 10^{-5}T-8.4\times 10^{-8}T^2~,\\
T &= &M_t /[\mbox{GeV}] -165~,\nn
\label{aem}
\ee
the matching condition (\ref{match}) and the GYU
boundary condition at $M_{\rm GUT}$ can be satisfied only for a specific
value of $\tan\beta$. Here  $M_{\tau},M_t, M_Z$
are pole masses, and the couplings above are defined in the 
$\overline{\mbox{MS}}$ scheme with six flavors.
Under the assumptions specified above, it is possible 
without knowing the
details of the scalar sector of the MSSM
to predict various 
parameters such as
 the top quark 
mass \cite{kmz1}-\cite{kmoz2}.
We present them for the model ${\bf A}$ in
table 1 and for the model ${\bf B}$ in
table 2.
 \begin{table}
\caption{Table 1: The predictions 
for different $M_{s}$ for ${\bf A}$}
\begin{center}
\begin{tabular}{|c|c|c|c|c|c|}
\hline
$M_{s}$ [GeV]   &$\alpha_{3(5{\rm f})}(M_Z)$ &
$\tan \beta$  &  $M_{\rm GUT}$ [GeV] 
 & $M_{b}$ [GeV]& $M_{t}$ [GeV]
\\ \hline
$300$ & $0.123 $  &$54.1 $  & $2.2\times 10^{16}$
 & $5.3$  & 183\\ \hline
$500$ & $0.122 $  &$54.2 $  & $1.9\times 10^{16}$
 & $5.3$  & 183 \\ \hline
$10^3$ & $0.120 $  &$54.3 $  & $1.5 \times 10^{16}$
 & $5.2$  & 184 \\ \hline
\end{tabular}
\end{center}
\end{table}

  \begin{table}
\caption{Table 2: The predictions 
for different $M_{s}$ for ${\bf B}$}
\begin{center}
\begin{tabular}{|c|c|c|c|c|c|}
\hline
$M_{s}$ [GeV]   &$\alpha_{3(5{\rm f})}(M_Z)$ &
$\tan \beta$  &  $M_{\rm GUT}$ [GeV] 
 & $M_{b}$ [GeV]& $M_{t}$ [GeV]
\\ \hline
$800$ & $0.120 $  &$48.2 $  & $1.5\times 10^{16}$
 & $5.4$  & 174\\ \hline
$10^3$ & $0.119 $  &$48.2 $  & $1.4\times 10^{16}$
 & $5.4$  & 174 \\ \hline
$1.2 \times 10^3$ & $0.118 $  &$48.2 $  & $1.3\times 10^{16}$
 & $5.4$  & 174 \\ \hline
\end{tabular}
\end{center}
\end{table}

\noindent
Comparing, for instance, 
the $M_t$ predictions above  
 with the most recent
experimental value  \cite{topmass}, 
\be
M_t = (175.6 \pm 5.5) ~~\mbox{GeV }~,
\ee
and recalling that
the theoretical values for $M_t$ given in the tables may suffer from
a correction  of less than  $\sim 4$ \% \cite{kmoz2},
we see that they are consistent with the experimental
data. 
(For more details, see ref. \cite{kmoz2},
where various corrections on the
predictions of GYU models such as 
the MSSM threshold corrections are estimated
\footnote{
The GUT threshold corrections in  the $SU(5)$ finite model are
given in ref, \cite{yoshioka1}.}.)

Now we come to the SSB sector.
As mentioned, we impose at  $M_{\rm GUT}$
 the $h=-MY$ relation (\ref{hY}) and
the soft scalar-mass sum rule (\ref{sumr}),
 i.e. (\ref{suma}) and (\ref{suma1}) for the models ${\bf A}$ and
(\ref{sumb}) and (\ref{sumb1})
for the model ${\bf B}$, 
 and calculate their low-energy values.
To make our unification idea and its consequence transparent,
we shall make an oversimplifying assumption
that the unique supersymmetry breaking scale
$M_{s}$
can be set equal to the unified gaugino mass $M$
at $M_{\rm GUT}$.
That is, we calculate the SSB parameters at 
$M_s=M$ from which we then compute the spectrum of the superpartners
by using the tree-level
formulae \footnote{For the lightest 
Higgs mass we include rediative 
corrections.}.
Since $\tan\beta$
 by the dimension-zero sector
because of GYU,
one should examine each time whether GYU 
and the sum rule are consistent with the radiative 
breaking of the electroweak symmetry \cite{inoue1}.
This concitency can be achieved, though not always,
by using the freedom to fix the $b$ term and the supersymmetric mass term
$\mu$ which remain unconstrained by finiteness.

As we can see from  (\ref{suma}) and (\ref{sumb}),
the structure of the sum rules for the two models is different.
Recall that  the MSSM Higgs
doublets,   $H_u$ and $H_d$,  mostly stem from 
the third Higgsess $H_3$ and $\overline{H}_3$
\footnote{For the model ${\bf A}$, this is an assumption
as we have discussed, while for ${\bf B}$ this is a consequence
of the unitarity of the mixing matrix of the three 
Higgsess \cite{kazakov1}.}.
Therefore, the scalar masses $m^{2}_{i}$ with
$i=H_1,H_2,\overline{H}_1, \overline{H}_2$ 
do not enter into the low-energy sector,
implying that $m^{2}_{{\bf 10}_1}~,~
m^{2}_{\overline{{\bf 5}}_1}~,~
m^{2}_{{\bf 10}_2}$ and $m^{2}_{\overline{{\bf 5}}_2}$
for the model ${\bf A}$, and 
$m^{2}_{{\bf 10}_1}$ and 
$m^{2}_{\overline{{\bf 5}}_1}$ for the ${\bf B}$,
respectively, are free parameters.
So in following discussions 
we would like to focus on the third-generation
scalar-masses. 
The  relevant sum rules 
at the GUT scale are thus given by
\be
m^{2}_{H_u}+
2  m^{2}_{{\bf 10}} &=&
m^{2}_{H_d}+ m^{2}_{\overline{{\bf 5}}}+
m^{2}_{{\bf 10}}=M^2~~\mbox{for}~~{\bf A} ~,\\
m^{2}_{H_u}+
2  m^{2}_{{\bf 10}} &=&M^2~,~
m^{2}_{H_d}-2m^{2}_{{\bf 10}}=-\frac{M^2}{3}~,~
m^{2}_{\overline{{\bf 5}}}+
3m^{2}_{{\bf 10}}=\frac{4M^2}{3}~~~\mbox{for}~~{\bf B},
\ee
where we use as  free parameters 
$m_{\overline{{\bf 5}}}\equiv m_{\overline{{\bf 5}}_3}$ and 
$m_{{\bf 10}}\equiv m_{{\bf 10}_3}$
for the model ${\bf A}$, and 
$m_{{\bf 10}}$ for ${\bf B}$, in addition to $M$.

First we present the result for the model  ${\bf A}$.
We look for the parameter space 
in which the lighter s-tau mass squared $m^2_{\tilde \tau}$
is larger than the lightest neutralino mass squared $m^2_\chi$
(which is the LSP).
In fig. 1, 2 and 3 we show this region in 
the $m_{\overline{{\bf 5}}}-m_{{\bf 10}}$
 plane for $M=M_{s} =0.3, 0.5$ and $1$ TeV, respectively.
The region with open squares does not lead to a successful 
radiative electroweak symmetry breaking, and the region with
dots and crosses defines the region with
 $m^2_{\tilde \tau} <0 $ and 
$m^2_{\tilde \tau} < m^2_\chi$, respectively.
\begin{center}
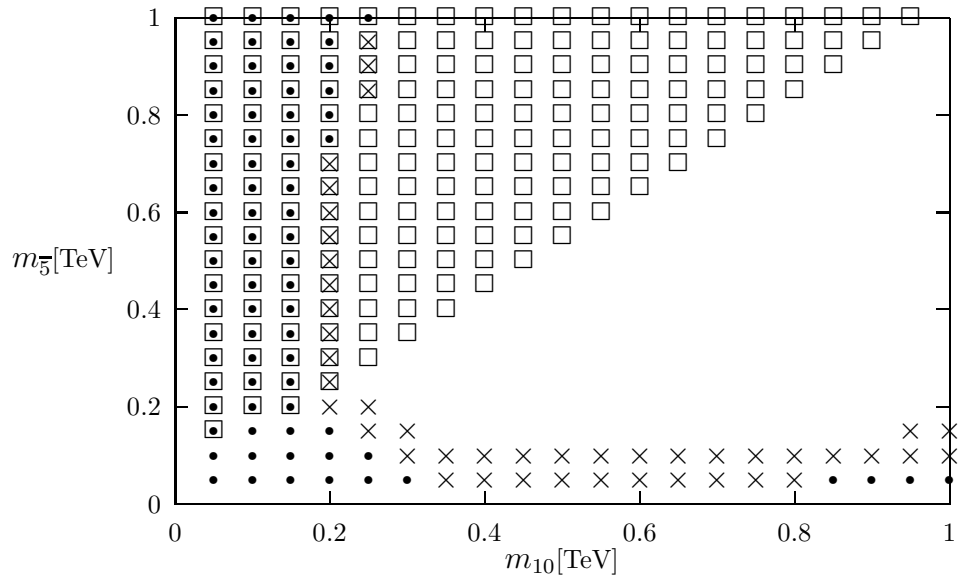
\input{figb3.tex} \\Fig. 1: The region without squares, dots and crosses
yields a neutralino  as the  LSP
 for the model ${\bf A}$  with $M=0.3$ TeV.
\end{center}
\begin{center}
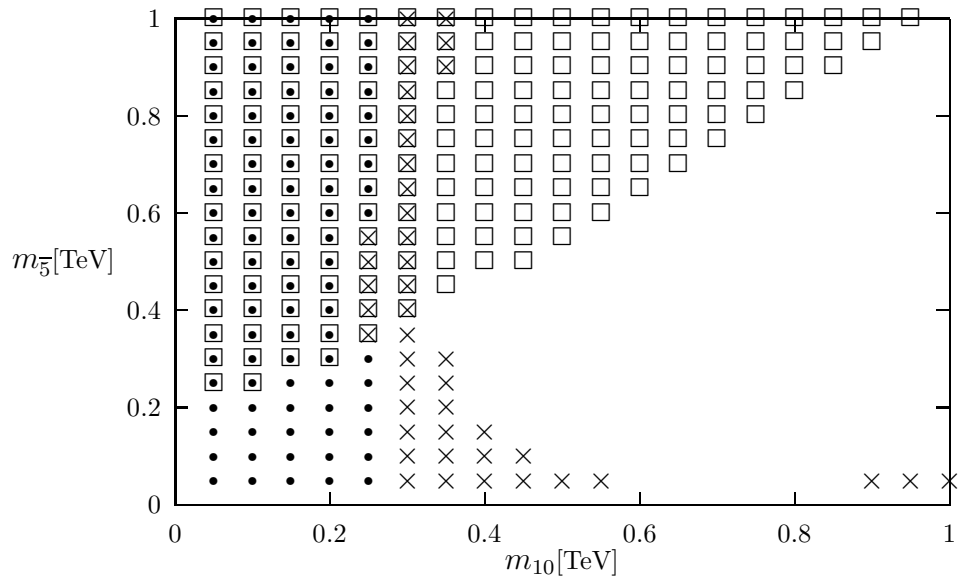
\input{figb5.tex} \\Fig. 2: The same as fig. 1  with $M=0.5$ TeV.
\end{center}
\begin{center}
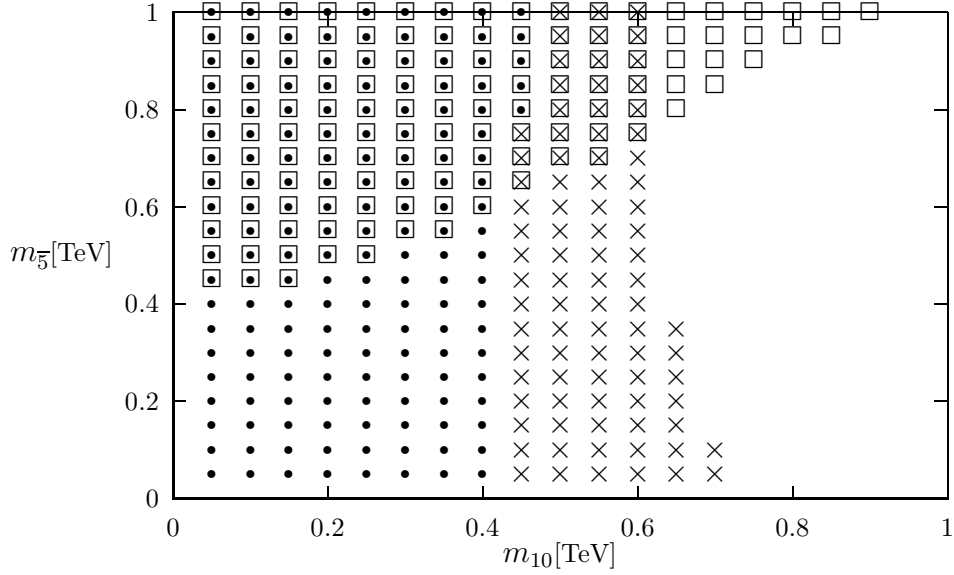
\input{figb10.tex} \\Fig. 3: The same as fig. 1  with $M=1$ TeV.
\end{center}
\begin{center}
\setlength{\unitlength}{0.240900pt}
\ifx\plotpoint\undefined\newsavebox{\plotpoint}\fi
\sbox{\plotpoint}{\rule[-0.200pt]{0.400pt}{0.400pt}}%
\begin{picture}(1500,900)(0,0)
\font\gnuplot=cmr10 at 10pt
\gnuplot
\sbox{\plotpoint}{\rule[-0.200pt]{0.400pt}{0.400pt}}%
\put(220.0,222.0){\rule[-0.200pt]{292.934pt}{0.400pt}}
\put(220.0,113.0){\rule[-0.200pt]{0.400pt}{184.048pt}}
\put(220.0,113.0){\rule[-0.200pt]{4.818pt}{0.400pt}}
\put(198,113){\makebox(0,0)[r]{-0.1}}
\put(1416.0,113.0){\rule[-0.200pt]{4.818pt}{0.400pt}}
\put(220.0,222.0){\rule[-0.200pt]{4.818pt}{0.400pt}}
\put(198,222){\makebox(0,0)[r]{0}}
\put(1416.0,222.0){\rule[-0.200pt]{4.818pt}{0.400pt}}
\put(220.0,331.0){\rule[-0.200pt]{4.818pt}{0.400pt}}
\put(198,331){\makebox(0,0)[r]{0.1}}
\put(1416.0,331.0){\rule[-0.200pt]{4.818pt}{0.400pt}}
\put(220.0,440.0){\rule[-0.200pt]{4.818pt}{0.400pt}}
\put(198,440){\makebox(0,0)[r]{0.2}}
\put(1416.0,440.0){\rule[-0.200pt]{4.818pt}{0.400pt}}
\put(220.0,550.0){\rule[-0.200pt]{4.818pt}{0.400pt}}
\put(198,550){\makebox(0,0)[r]{0.3}}
\put(1416.0,550.0){\rule[-0.200pt]{4.818pt}{0.400pt}}
\put(220.0,659.0){\rule[-0.200pt]{4.818pt}{0.400pt}}
\put(198,659){\makebox(0,0)[r]{0.4}}
\put(1416.0,659.0){\rule[-0.200pt]{4.818pt}{0.400pt}}
\put(220.0,768.0){\rule[-0.200pt]{4.818pt}{0.400pt}}
\put(198,768){\makebox(0,0)[r]{0.5}}
\put(1416.0,768.0){\rule[-0.200pt]{4.818pt}{0.400pt}}
\put(220.0,877.0){\rule[-0.200pt]{4.818pt}{0.400pt}}
\put(198,877){\makebox(0,0)[r]{0.6}}
\put(1416.0,877.0){\rule[-0.200pt]{4.818pt}{0.400pt}}
\put(220.0,113.0){\rule[-0.200pt]{0.400pt}{4.818pt}}
\put(220,68){\makebox(0,0){0}}
\put(220.0,857.0){\rule[-0.200pt]{0.400pt}{4.818pt}}
\put(372.0,113.0){\rule[-0.200pt]{0.400pt}{4.818pt}}
\put(372,68){\makebox(0,0){200}}
\put(372.0,857.0){\rule[-0.200pt]{0.400pt}{4.818pt}}
\put(524.0,113.0){\rule[-0.200pt]{0.400pt}{4.818pt}}
\put(524,68){\makebox(0,0){400}}
\put(524.0,857.0){\rule[-0.200pt]{0.400pt}{4.818pt}}
\put(676.0,113.0){\rule[-0.200pt]{0.400pt}{4.818pt}}
\put(676,68){\makebox(0,0){600}}
\put(676.0,857.0){\rule[-0.200pt]{0.400pt}{4.818pt}}
\put(828.0,113.0){\rule[-0.200pt]{0.400pt}{4.818pt}}
\put(828,68){\makebox(0,0){800}}
\put(828.0,857.0){\rule[-0.200pt]{0.400pt}{4.818pt}}
\put(980.0,113.0){\rule[-0.200pt]{0.400pt}{4.818pt}}
\put(980,68){\makebox(0,0){1000}}
\put(980.0,857.0){\rule[-0.200pt]{0.400pt}{4.818pt}}
\put(1132.0,113.0){\rule[-0.200pt]{0.400pt}{4.818pt}}
\put(1132,68){\makebox(0,0){1200}}
\put(1132.0,857.0){\rule[-0.200pt]{0.400pt}{4.818pt}}
\put(1284.0,113.0){\rule[-0.200pt]{0.400pt}{4.818pt}}
\put(1284,68){\makebox(0,0){1400}}
\put(1284.0,857.0){\rule[-0.200pt]{0.400pt}{4.818pt}}
\put(1436.0,113.0){\rule[-0.200pt]{0.400pt}{4.818pt}}
\put(1436,68){\makebox(0,0){1600}}
\put(1436.0,857.0){\rule[-0.200pt]{0.400pt}{4.818pt}}
\put(220.0,113.0){\rule[-0.200pt]{292.934pt}{0.400pt}}
\put(1436.0,113.0){\rule[-0.200pt]{0.400pt}{184.048pt}}
\put(220.0,877.0){\rule[-0.200pt]{292.934pt}{0.400pt}}
\put(45,495){\makebox(0,0){[TeV]$^2$}}
\put(828,23){\makebox(0,0){$M$[GeV]}}
\put(1056,550){\makebox(0,0)[r]{$m^2_\chi$}}
\put(1208,386){\makebox(0,0)[r]{$m^2_{\tilde \tau}$}}
\put(220.0,113.0){\rule[-0.200pt]{0.400pt}{184.048pt}}
\put(296,224){\usebox{\plotpoint}}
\multiput(296.00,224.59)(8.391,0.477){7}{\rule{6.180pt}{0.115pt}}
\multiput(296.00,223.17)(63.173,5.000){2}{\rule{3.090pt}{0.400pt}}
\multiput(372.00,229.58)(3.935,0.491){17}{\rule{3.140pt}{0.118pt}}
\multiput(372.00,228.17)(69.483,10.000){2}{\rule{1.570pt}{0.400pt}}
\multiput(448.00,239.58)(2.584,0.494){27}{\rule{2.127pt}{0.119pt}}
\multiput(448.00,238.17)(71.586,15.000){2}{\rule{1.063pt}{0.400pt}}
\multiput(524.00,254.58)(2.143,0.495){33}{\rule{1.789pt}{0.119pt}}
\multiput(524.00,253.17)(72.287,18.000){2}{\rule{0.894pt}{0.400pt}}
\multiput(600.00,272.58)(1.598,0.496){45}{\rule{1.367pt}{0.120pt}}
\multiput(600.00,271.17)(73.163,24.000){2}{\rule{0.683pt}{0.400pt}}
\multiput(676.00,296.58)(1.367,0.497){53}{\rule{1.186pt}{0.120pt}}
\multiput(676.00,295.17)(73.539,28.000){2}{\rule{0.593pt}{0.400pt}}
\multiput(752.00,324.58)(1.157,0.497){63}{\rule{1.021pt}{0.120pt}}
\multiput(752.00,323.17)(73.880,33.000){2}{\rule{0.511pt}{0.400pt}}
\multiput(828.00,357.58)(1.031,0.498){71}{\rule{0.922pt}{0.120pt}}
\multiput(828.00,356.17)(74.087,37.000){2}{\rule{0.461pt}{0.400pt}}
\multiput(904.00,394.58)(0.907,0.498){81}{\rule{0.824pt}{0.120pt}}
\multiput(904.00,393.17)(74.290,42.000){2}{\rule{0.412pt}{0.400pt}}
\multiput(980.00,436.58)(0.810,0.498){91}{\rule{0.747pt}{0.120pt}}
\multiput(980.00,435.17)(74.450,47.000){2}{\rule{0.373pt}{0.400pt}}
\multiput(1056.00,483.58)(0.732,0.498){101}{\rule{0.685pt}{0.120pt}}
\multiput(1056.00,482.17)(74.579,52.000){2}{\rule{0.342pt}{0.400pt}}
\multiput(1132.00,535.58)(0.667,0.499){111}{\rule{0.633pt}{0.120pt}}
\multiput(1132.00,534.17)(74.685,57.000){2}{\rule{0.317pt}{0.400pt}}
\multiput(1208.00,592.58)(0.613,0.499){121}{\rule{0.590pt}{0.120pt}}
\multiput(1208.00,591.17)(74.775,62.000){2}{\rule{0.295pt}{0.400pt}}
\multiput(1284.00,654.58)(0.567,0.499){131}{\rule{0.554pt}{0.120pt}}
\multiput(1284.00,653.17)(74.851,67.000){2}{\rule{0.277pt}{0.400pt}}
\multiput(1360.00,721.58)(0.535,0.499){139}{\rule{0.528pt}{0.120pt}}
\multiput(1360.00,720.17)(74.904,71.000){2}{\rule{0.264pt}{0.400pt}}
\sbox{\plotpoint}{\rule[-0.400pt]{0.800pt}{0.800pt}}%
\put(296,212){\usebox{\plotpoint}}
\put(296,208.84){\rule{18.308pt}{0.800pt}}
\multiput(296.00,210.34)(38.000,-3.000){2}{\rule{9.154pt}{0.800pt}}
\put(372,209.34){\rule{15.400pt}{0.800pt}}
\multiput(372.00,207.34)(44.037,4.000){2}{\rule{7.700pt}{0.800pt}}
\multiput(448.00,214.40)(3.730,0.512){15}{\rule{5.727pt}{0.123pt}}
\multiput(448.00,211.34)(64.113,11.000){2}{\rule{2.864pt}{0.800pt}}
\multiput(524.00,225.41)(2.652,0.508){23}{\rule{4.253pt}{0.122pt}}
\multiput(524.00,222.34)(67.172,15.000){2}{\rule{2.127pt}{0.800pt}}
\multiput(600.00,240.41)(2.063,0.506){31}{\rule{3.400pt}{0.122pt}}
\multiput(600.00,237.34)(68.943,19.000){2}{\rule{1.700pt}{0.800pt}}
\multiput(676.00,259.41)(1.769,0.505){37}{\rule{2.964pt}{0.122pt}}
\multiput(676.00,256.34)(69.849,22.000){2}{\rule{1.482pt}{0.800pt}}
\multiput(752.00,281.41)(1.488,0.504){45}{\rule{2.538pt}{0.121pt}}
\multiput(752.00,278.34)(70.731,26.000){2}{\rule{1.269pt}{0.800pt}}
\multiput(828.00,307.41)(1.284,0.503){53}{\rule{2.227pt}{0.121pt}}
\multiput(828.00,304.34)(71.378,30.000){2}{\rule{1.113pt}{0.800pt}}
\multiput(904.00,337.41)(1.201,0.503){57}{\rule{2.100pt}{0.121pt}}
\multiput(904.00,334.34)(71.641,32.000){2}{\rule{1.050pt}{0.800pt}}
\multiput(980.00,369.41)(1.065,0.503){65}{\rule{1.889pt}{0.121pt}}
\multiput(980.00,366.34)(72.080,36.000){2}{\rule{0.944pt}{0.800pt}}
\multiput(1056.00,405.41)(0.981,0.503){71}{\rule{1.759pt}{0.121pt}}
\multiput(1056.00,402.34)(72.349,39.000){2}{\rule{0.879pt}{0.800pt}}
\multiput(1132.00,444.41)(0.910,0.502){77}{\rule{1.648pt}{0.121pt}}
\multiput(1132.00,441.34)(72.580,42.000){2}{\rule{0.824pt}{0.800pt}}
\multiput(1208.00,486.41)(0.848,0.502){83}{\rule{1.551pt}{0.121pt}}
\multiput(1208.00,483.34)(72.781,45.000){2}{\rule{0.776pt}{0.800pt}}
\multiput(1284.00,531.41)(0.794,0.502){89}{\rule{1.467pt}{0.121pt}}
\multiput(1284.00,528.34)(72.956,48.000){2}{\rule{0.733pt}{0.800pt}}
\multiput(1360.00,579.41)(0.732,0.502){97}{\rule{1.369pt}{0.121pt}}
\multiput(1360.00,576.34)(73.158,52.000){2}{\rule{0.685pt}{0.800pt}}
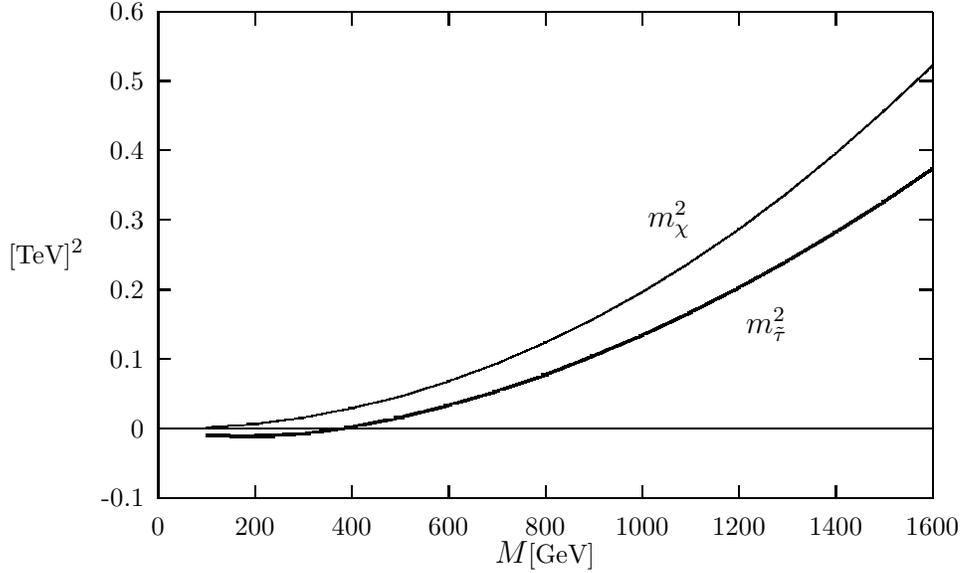
\end{picture} \\Fig. 4: $m^2_{\tilde \tau}$ and  $m^2_\chi$
for the universal choice of the soft scalar masses.
\end{center}
In fig. 4 we show 
$m^2_{\tilde \tau}$ and  $m^2_\chi$
for the universal choice
$m^2_{\tilde \tau}= m^2_\chi=M^2 /3$ at $M_{\rm GUT}$.
We find  that there is no region of $M_s=M$ 
below $O$(few) TeV
in which $m^2_{\tilde \tau} > m^2_\chi$
is satisfied.
In table 3 we present  the s-spectrum 
and the lightest Higgs mass $m_h$ of the model
${\bf A}$ with  $M=0.5$ TeV, 
 $m_{\overline{{\bf 5}}} =0.3$ TeV and 
$m_{{\bf 10}}=0.5$ TeV.
(Radiative corrections are included in $m_h$.)

\begin{center}
Table 3:  A representative
example of the predictions for the s-spectrum
for the model ${\bf A}$.

\vspace{0.2cm}
\begin{tabular}{|c|c||c|c|}
 \hline
$m_{\chi}=m_{\chi_1}$ (TeV) &
0.22 &
$m_{\tilde{b}_2}$ (TeV)  &
1.06 \\ \hline
$m_{\chi_2}$ (TeV) &
0.41 &
$m_{\tilde{\tau}}=m_{\tilde{\tau}_1}$ (TeV)  &
0.33
\\ \hline
$m_{\chi_3} $ (TeV) &
0.93 &
$m_{\tilde{\tau}_2}$ (TeV)   &
0.54
\\ \hline
$m_{\chi_4}$ (TeV) & 
0.94 &
$m_{\tilde{\nu}_1}$ (TeV)   &
0.41
\\ \hline
$m_{\chi^{\pm}_{1}} $ (TeV) &
0.41 &
$m_{A}$ (TeV) &
0.44
\\ \hline
$m_{\chi^{\pm}_{2}} $ (TeV) &
0.94 &
$ m_{H^{\pm}}$ 
(TeV) & 
0.45
\\ \hline
$m_{\tilde{t}_1}$ (TeV) &
0.94 &
$ m_{H}$ 
(TeV) &
0.44
\\ \hline
$m_{\tilde{t}_2}$  (TeV) &
1.09 &
$m_{h} $
(TeV) & 0.12
\\ \hline
$m_{\tilde{b}_1}$  (TeV) &
0.86  & & 
\\  \hline
\end{tabular}

\end{center}

\vspace{0.2cm}
The model ${\bf B}$ has only two free SSB parameters
$ m_{{\bf 10}}$ and $M=(M_s)$.
For a fixed $M$, the neutralino masses are independent
of $m_{{\bf 10}}$, while
$m_{\tilde \tau}$ depends on it.
Shown are $m^2_{\tilde \tau}$ and $m^2_\chi$
as function of $ m_{{\bf 10}}$ 
 in fig. 5, 6 and 7
for $M=0.5, 0.8$ and $1$ TeV. 
\begin{center}
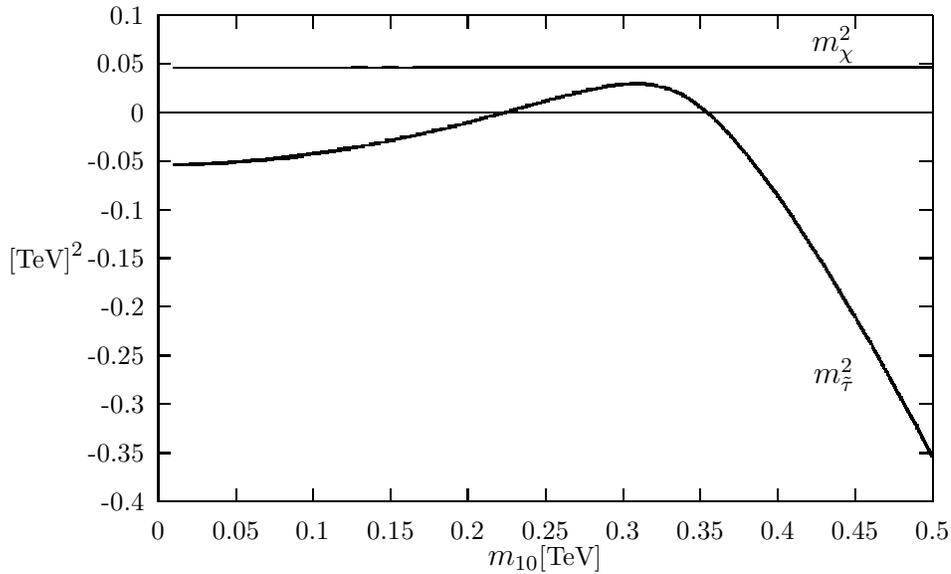
\input{fig7.tex} \\Fig. 5:  $m^2_{\tilde \tau}$
and $m^2_\chi$ against $m^{2}_{{\bf 10}}$
for  $M=0.5$ TeV. 
\end{center}
\begin{center}
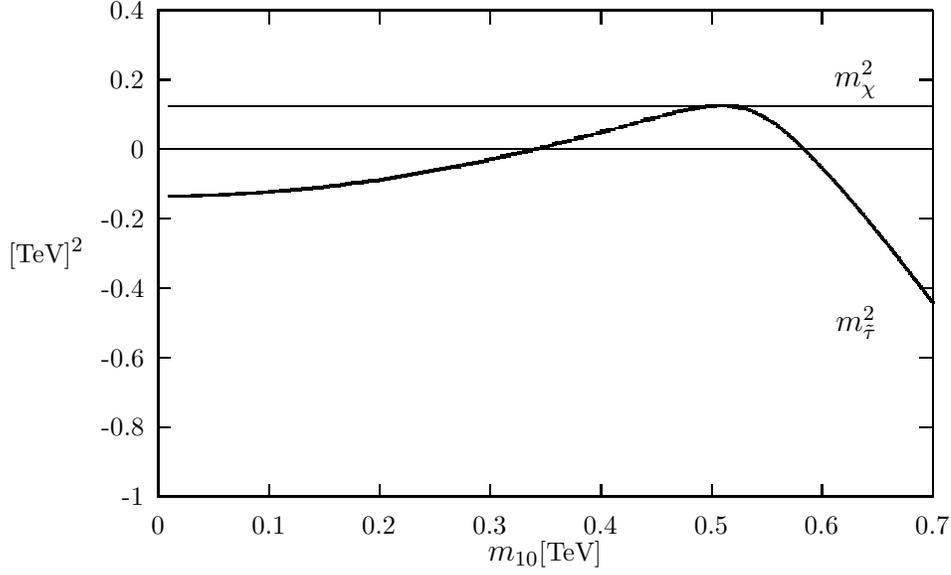
\input{fig8.tex} \\Fig. 6:  The same as fig. 4 for  $M=0.8$ TeV. 
\end{center}
\begin{center}
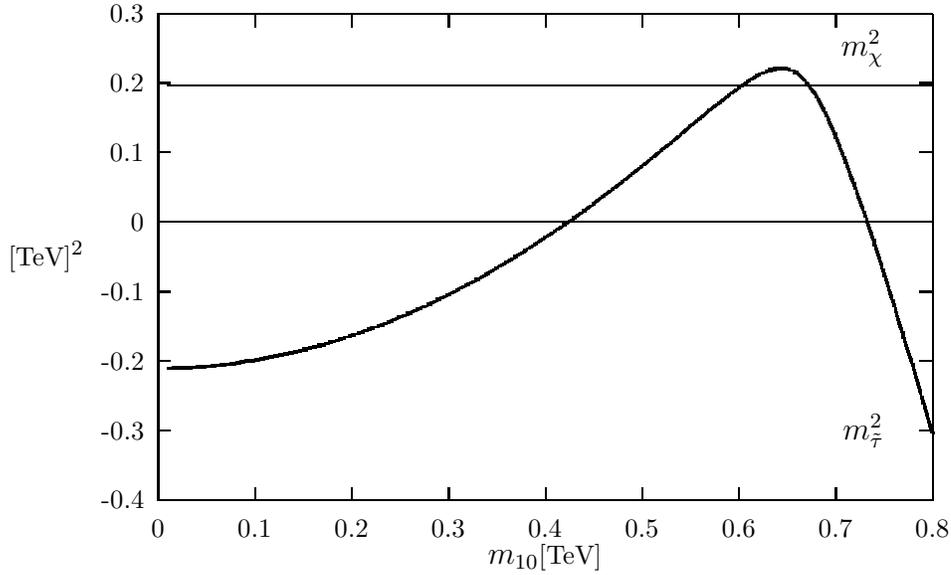
\input{fig9.tex} \\Fig. 7:  The same as fig. 4 for  $M=1$ TeV. 
\end{center}

In fig. 8 we plot 
 the maximal value of $m^2_{\tilde \tau}$,
denoted by  Max($m^2_{\tilde \tau}$),
and $m^2_\chi$ 
for  different values of $M$,
which should be compared with fig. 9 in which 
we plot the case of the universal choice of
the scalar masses.
\begin{center}
\setlength{\unitlength}{0.240900pt}
\ifx\plotpoint\undefined\newsavebox{\plotpoint}\fi
\sbox{\plotpoint}{\rule[-0.200pt]{0.400pt}{0.400pt}}%
\begin{picture}(1500,900)(0,0)
\font\gnuplot=cmr10 at 10pt
\gnuplot
\sbox{\plotpoint}{\rule[-0.200pt]{0.400pt}{0.400pt}}%
\put(220.0,189.0){\rule[-0.200pt]{292.934pt}{0.400pt}}
\put(220.0,113.0){\rule[-0.200pt]{0.400pt}{184.048pt}}
\put(220.0,113.0){\rule[-0.200pt]{4.818pt}{0.400pt}}
\put(198,113){\makebox(0,0)[r]{-0.05}}
\put(1416.0,113.0){\rule[-0.200pt]{4.818pt}{0.400pt}}
\put(220.0,189.0){\rule[-0.200pt]{4.818pt}{0.400pt}}
\put(198,189){\makebox(0,0)[r]{0}}
\put(1416.0,189.0){\rule[-0.200pt]{4.818pt}{0.400pt}}
\put(220.0,266.0){\rule[-0.200pt]{4.818pt}{0.400pt}}
\put(198,266){\makebox(0,0)[r]{0.05}}
\put(1416.0,266.0){\rule[-0.200pt]{4.818pt}{0.400pt}}
\put(220.0,342.0){\rule[-0.200pt]{4.818pt}{0.400pt}}
\put(198,342){\makebox(0,0)[r]{0.1}}
\put(1416.0,342.0){\rule[-0.200pt]{4.818pt}{0.400pt}}
\put(220.0,419.0){\rule[-0.200pt]{4.818pt}{0.400pt}}
\put(198,419){\makebox(0,0)[r]{0.15}}
\put(1416.0,419.0){\rule[-0.200pt]{4.818pt}{0.400pt}}
\put(220.0,495.0){\rule[-0.200pt]{4.818pt}{0.400pt}}
\put(198,495){\makebox(0,0)[r]{0.2}}
\put(1416.0,495.0){\rule[-0.200pt]{4.818pt}{0.400pt}}
\put(220.0,571.0){\rule[-0.200pt]{4.818pt}{0.400pt}}
\put(198,571){\makebox(0,0)[r]{0.25}}
\put(1416.0,571.0){\rule[-0.200pt]{4.818pt}{0.400pt}}
\put(220.0,648.0){\rule[-0.200pt]{4.818pt}{0.400pt}}
\put(198,648){\makebox(0,0)[r]{0.3}}
\put(1416.0,648.0){\rule[-0.200pt]{4.818pt}{0.400pt}}
\put(220.0,724.0){\rule[-0.200pt]{4.818pt}{0.400pt}}
\put(198,724){\makebox(0,0)[r]{0.35}}
\put(1416.0,724.0){\rule[-0.200pt]{4.818pt}{0.400pt}}
\put(220.0,801.0){\rule[-0.200pt]{4.818pt}{0.400pt}}
\put(198,801){\makebox(0,0)[r]{0.4}}
\put(1416.0,801.0){\rule[-0.200pt]{4.818pt}{0.400pt}}
\put(220.0,877.0){\rule[-0.200pt]{4.818pt}{0.400pt}}
\put(198,877){\makebox(0,0)[r]{0.45}}
\put(1416.0,877.0){\rule[-0.200pt]{4.818pt}{0.400pt}}
\put(220.0,113.0){\rule[-0.200pt]{0.400pt}{4.818pt}}
\put(220,68){\makebox(0,0){0}}
\put(220.0,857.0){\rule[-0.200pt]{0.400pt}{4.818pt}}
\put(394.0,113.0){\rule[-0.200pt]{0.400pt}{4.818pt}}
\put(394,68){\makebox(0,0){200}}
\put(394.0,857.0){\rule[-0.200pt]{0.400pt}{4.818pt}}
\put(567.0,113.0){\rule[-0.200pt]{0.400pt}{4.818pt}}
\put(567,68){\makebox(0,0){400}}
\put(567.0,857.0){\rule[-0.200pt]{0.400pt}{4.818pt}}
\put(741.0,113.0){\rule[-0.200pt]{0.400pt}{4.818pt}}
\put(741,68){\makebox(0,0){600}}
\put(741.0,857.0){\rule[-0.200pt]{0.400pt}{4.818pt}}
\put(915.0,113.0){\rule[-0.200pt]{0.400pt}{4.818pt}}
\put(915,68){\makebox(0,0){800}}
\put(915.0,857.0){\rule[-0.200pt]{0.400pt}{4.818pt}}
\put(1089.0,113.0){\rule[-0.200pt]{0.400pt}{4.818pt}}
\put(1089,68){\makebox(0,0){1000}}
\put(1089.0,857.0){\rule[-0.200pt]{0.400pt}{4.818pt}}
\put(1262.0,113.0){\rule[-0.200pt]{0.400pt}{4.818pt}}
\put(1262,68){\makebox(0,0){1200}}
\put(1262.0,857.0){\rule[-0.200pt]{0.400pt}{4.818pt}}
\put(1436.0,113.0){\rule[-0.200pt]{0.400pt}{4.818pt}}
\put(1436,68){\makebox(0,0){1400}}
\put(1436.0,857.0){\rule[-0.200pt]{0.400pt}{4.818pt}}
\put(220.0,113.0){\rule[-0.200pt]{292.934pt}{0.400pt}}
\put(1436.0,113.0){\rule[-0.200pt]{0.400pt}{184.048pt}}
\put(220.0,877.0){\rule[-0.200pt]{292.934pt}{0.400pt}}
\put(45,495){\makebox(0,0){[TeV]$^2$}}
\put(828,23){\makebox(0,0){$M$[GeV]}}
\put(1349,571){\makebox(0,0)[r]{$m^2_\chi$}}
\put(1175,648){\makebox(0,0)[r]{Max($m^2_{\tilde \tau})$}}
\put(220.0,113.0){\rule[-0.200pt]{0.400pt}{184.048pt}}
\put(307,191){\usebox{\plotpoint}}
\multiput(307.00,191.59)(5.711,0.488){13}{\rule{4.450pt}{0.117pt}}
\multiput(307.00,190.17)(77.764,8.000){2}{\rule{2.225pt}{0.400pt}}
\multiput(394.00,199.58)(3.178,0.494){25}{\rule{2.586pt}{0.119pt}}
\multiput(394.00,198.17)(81.633,14.000){2}{\rule{1.293pt}{0.400pt}}
\multiput(481.00,213.58)(2.073,0.496){39}{\rule{1.738pt}{0.119pt}}
\multiput(481.00,212.17)(82.392,21.000){2}{\rule{0.869pt}{0.400pt}}
\multiput(567.00,234.58)(1.688,0.497){49}{\rule{1.438pt}{0.120pt}}
\multiput(567.00,233.17)(84.014,26.000){2}{\rule{0.719pt}{0.400pt}}
\multiput(654.00,260.58)(1.326,0.497){63}{\rule{1.155pt}{0.120pt}}
\multiput(654.00,259.17)(84.604,33.000){2}{\rule{0.577pt}{0.400pt}}
\multiput(741.00,293.58)(1.120,0.498){75}{\rule{0.992pt}{0.120pt}}
\multiput(741.00,292.17)(84.940,39.000){2}{\rule{0.496pt}{0.400pt}}
\multiput(828.00,332.58)(0.948,0.498){89}{\rule{0.857pt}{0.120pt}}
\multiput(828.00,331.17)(85.222,46.000){2}{\rule{0.428pt}{0.400pt}}
\multiput(915.00,378.58)(0.822,0.498){103}{\rule{0.757pt}{0.120pt}}
\multiput(915.00,377.17)(85.430,53.000){2}{\rule{0.378pt}{0.400pt}}
\multiput(1002.00,431.58)(0.738,0.499){115}{\rule{0.690pt}{0.120pt}}
\multiput(1002.00,430.17)(85.568,59.000){2}{\rule{0.345pt}{0.400pt}}
\multiput(1089.00,490.58)(0.652,0.499){129}{\rule{0.621pt}{0.120pt}}
\multiput(1089.00,489.17)(84.711,66.000){2}{\rule{0.311pt}{0.400pt}}
\multiput(1175.00,556.58)(0.596,0.499){143}{\rule{0.577pt}{0.120pt}}
\multiput(1175.00,555.17)(85.803,73.000){2}{\rule{0.288pt}{0.400pt}}
\multiput(1262.00,629.58)(0.543,0.499){157}{\rule{0.535pt}{0.120pt}}
\multiput(1262.00,628.17)(85.890,80.000){2}{\rule{0.268pt}{0.400pt}}
\sbox{\plotpoint}{\rule[-0.400pt]{0.800pt}{0.800pt}}%
\put(307,180){\usebox{\plotpoint}}
\put(307,176.84){\rule{20.958pt}{0.800pt}}
\multiput(307.00,178.34)(43.500,-3.000){2}{\rule{10.479pt}{0.800pt}}
\multiput(394.00,178.40)(5.407,0.516){11}{\rule{7.933pt}{0.124pt}}
\multiput(394.00,175.34)(70.534,9.000){2}{\rule{3.967pt}{0.800pt}}
\multiput(481.00,187.41)(2.338,0.506){31}{\rule{3.821pt}{0.122pt}}
\multiput(481.00,184.34)(78.069,19.000){2}{\rule{1.911pt}{0.800pt}}
\multiput(567.00,206.41)(1.472,0.503){53}{\rule{2.520pt}{0.121pt}}
\multiput(567.00,203.34)(81.770,30.000){2}{\rule{1.260pt}{0.800pt}}
\multiput(654.00,236.41)(1.125,0.503){71}{\rule{1.985pt}{0.121pt}}
\multiput(654.00,233.34)(82.881,39.000){2}{\rule{0.992pt}{0.800pt}}
\multiput(741.00,275.41)(0.892,0.502){91}{\rule{1.620pt}{0.121pt}}
\multiput(741.00,272.34)(83.637,49.000){2}{\rule{0.810pt}{0.800pt}}
\multiput(828.00,324.41)(0.739,0.502){111}{\rule{1.380pt}{0.121pt}}
\multiput(828.00,321.34)(84.136,59.000){2}{\rule{0.690pt}{0.800pt}}
\multiput(915.00,383.41)(0.640,0.501){129}{\rule{1.224pt}{0.121pt}}
\multiput(915.00,380.34)(84.461,68.000){2}{\rule{0.612pt}{0.800pt}}
\multiput(1002.00,451.41)(0.565,0.501){147}{\rule{1.104pt}{0.121pt}}
\multiput(1002.00,448.34)(84.709,77.000){2}{\rule{0.552pt}{0.800pt}}
\multiput(1090.41,527.00)(0.501,0.511){165}{\rule{0.121pt}{1.019pt}}
\multiput(1087.34,527.00)(86.000,85.886){2}{\rule{0.800pt}{0.509pt}}
\multiput(1176.41,615.00)(0.501,0.551){167}{\rule{0.121pt}{1.083pt}}
\multiput(1173.34,615.00)(87.000,93.753){2}{\rule{0.800pt}{0.541pt}}
\multiput(1263.41,711.00)(0.501,0.609){167}{\rule{0.121pt}{1.175pt}}
\multiput(1260.34,711.00)(87.000,103.562){2}{\rule{0.800pt}{0.587pt}}
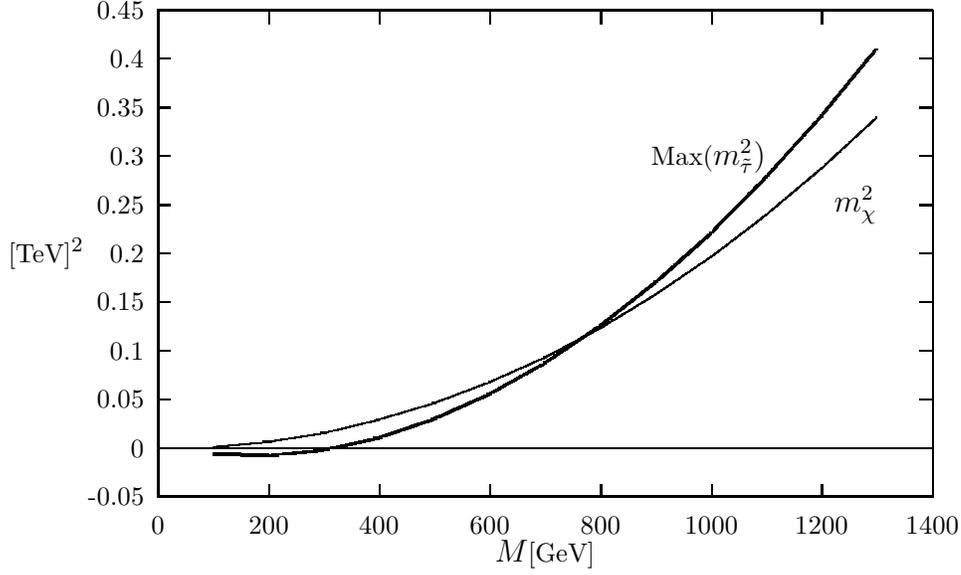
\end{picture} \\Fig. 8:  Max($m^2_{\tilde \tau}$)
and $m^2_\chi$ as function of $M$
\end{center}
\begin{center}
\setlength{\unitlength}{0.240900pt}
\ifx\plotpoint\undefined\newsavebox{\plotpoint}\fi
\sbox{\plotpoint}{\rule[-0.200pt]{0.400pt}{0.400pt}}%
\begin{picture}(1500,900)(0,0)
\font\gnuplot=cmr10 at 10pt
\gnuplot
\sbox{\plotpoint}{\rule[-0.200pt]{0.400pt}{0.400pt}}%
\put(220.0,209.0){\rule[-0.200pt]{292.934pt}{0.400pt}}
\put(220.0,113.0){\rule[-0.200pt]{0.400pt}{184.048pt}}
\put(220.0,113.0){\rule[-0.200pt]{4.818pt}{0.400pt}}
\put(198,113){\makebox(0,0)[r]{-0.05}}
\put(1416.0,113.0){\rule[-0.200pt]{4.818pt}{0.400pt}}
\put(220.0,209.0){\rule[-0.200pt]{4.818pt}{0.400pt}}
\put(198,209){\makebox(0,0)[r]{0}}
\put(1416.0,209.0){\rule[-0.200pt]{4.818pt}{0.400pt}}
\put(220.0,304.0){\rule[-0.200pt]{4.818pt}{0.400pt}}
\put(198,304){\makebox(0,0)[r]{0.05}}
\put(1416.0,304.0){\rule[-0.200pt]{4.818pt}{0.400pt}}
\put(220.0,400.0){\rule[-0.200pt]{4.818pt}{0.400pt}}
\put(198,400){\makebox(0,0)[r]{0.1}}
\put(1416.0,400.0){\rule[-0.200pt]{4.818pt}{0.400pt}}
\put(220.0,495.0){\rule[-0.200pt]{4.818pt}{0.400pt}}
\put(198,495){\makebox(0,0)[r]{0.15}}
\put(1416.0,495.0){\rule[-0.200pt]{4.818pt}{0.400pt}}
\put(220.0,591.0){\rule[-0.200pt]{4.818pt}{0.400pt}}
\put(198,591){\makebox(0,0)[r]{0.2}}
\put(1416.0,591.0){\rule[-0.200pt]{4.818pt}{0.400pt}}
\put(220.0,686.0){\rule[-0.200pt]{4.818pt}{0.400pt}}
\put(198,686){\makebox(0,0)[r]{0.25}}
\put(1416.0,686.0){\rule[-0.200pt]{4.818pt}{0.400pt}}
\put(220.0,782.0){\rule[-0.200pt]{4.818pt}{0.400pt}}
\put(198,782){\makebox(0,0)[r]{0.3}}
\put(1416.0,782.0){\rule[-0.200pt]{4.818pt}{0.400pt}}
\put(220.0,877.0){\rule[-0.200pt]{4.818pt}{0.400pt}}
\put(198,877){\makebox(0,0)[r]{0.35}}
\put(1416.0,877.0){\rule[-0.200pt]{4.818pt}{0.400pt}}
\put(220.0,113.0){\rule[-0.200pt]{0.400pt}{4.818pt}}
\put(220,68){\makebox(0,0){0}}
\put(220.0,857.0){\rule[-0.200pt]{0.400pt}{4.818pt}}
\put(394.0,113.0){\rule[-0.200pt]{0.400pt}{4.818pt}}
\put(394,68){\makebox(0,0){200}}
\put(394.0,857.0){\rule[-0.200pt]{0.400pt}{4.818pt}}
\put(567.0,113.0){\rule[-0.200pt]{0.400pt}{4.818pt}}
\put(567,68){\makebox(0,0){400}}
\put(567.0,857.0){\rule[-0.200pt]{0.400pt}{4.818pt}}
\put(741.0,113.0){\rule[-0.200pt]{0.400pt}{4.818pt}}
\put(741,68){\makebox(0,0){600}}
\put(741.0,857.0){\rule[-0.200pt]{0.400pt}{4.818pt}}
\put(915.0,113.0){\rule[-0.200pt]{0.400pt}{4.818pt}}
\put(915,68){\makebox(0,0){800}}
\put(915.0,857.0){\rule[-0.200pt]{0.400pt}{4.818pt}}
\put(1089.0,113.0){\rule[-0.200pt]{0.400pt}{4.818pt}}
\put(1089,68){\makebox(0,0){1000}}
\put(1089.0,857.0){\rule[-0.200pt]{0.400pt}{4.818pt}}
\put(1262.0,113.0){\rule[-0.200pt]{0.400pt}{4.818pt}}
\put(1262,68){\makebox(0,0){1200}}
\put(1262.0,857.0){\rule[-0.200pt]{0.400pt}{4.818pt}}
\put(1436.0,113.0){\rule[-0.200pt]{0.400pt}{4.818pt}}
\put(1436,68){\makebox(0,0){1400}}
\put(1436.0,857.0){\rule[-0.200pt]{0.400pt}{4.818pt}}
\put(220.0,113.0){\rule[-0.200pt]{292.934pt}{0.400pt}}
\put(1436.0,113.0){\rule[-0.200pt]{0.400pt}{184.048pt}}
\put(220.0,877.0){\rule[-0.200pt]{292.934pt}{0.400pt}}
\put(45,495){\makebox(0,0){[TeV]$^2$}}
\put(828,23){\makebox(0,0){$M$[GeV]}}
\put(1175,782){\makebox(0,0)[r]{$m^2_\chi$}}
\put(1349,591){\makebox(0,0)[r]{$m^2_{\tilde \tau}$}}
\put(220.0,113.0){\rule[-0.200pt]{0.400pt}{184.048pt}}
\put(307,211){\usebox{\plotpoint}}
\multiput(307.00,211.58)(4.509,0.491){17}{\rule{3.580pt}{0.118pt}}
\multiput(307.00,210.17)(79.570,10.000){2}{\rule{1.790pt}{0.400pt}}
\multiput(394.00,221.58)(2.455,0.495){33}{\rule{2.033pt}{0.119pt}}
\multiput(394.00,220.17)(82.780,18.000){2}{\rule{1.017pt}{0.400pt}}
\multiput(481.00,239.58)(1.736,0.497){47}{\rule{1.476pt}{0.120pt}}
\multiput(481.00,238.17)(82.936,25.000){2}{\rule{0.738pt}{0.400pt}}
\multiput(567.00,264.58)(1.326,0.497){63}{\rule{1.155pt}{0.120pt}}
\multiput(567.00,263.17)(84.604,33.000){2}{\rule{0.577pt}{0.400pt}}
\multiput(654.00,297.58)(1.065,0.498){79}{\rule{0.949pt}{0.120pt}}
\multiput(654.00,296.17)(85.031,41.000){2}{\rule{0.474pt}{0.400pt}}
\multiput(741.00,338.58)(0.890,0.498){95}{\rule{0.810pt}{0.120pt}}
\multiput(741.00,337.17)(85.318,49.000){2}{\rule{0.405pt}{0.400pt}}
\multiput(828.00,387.58)(0.764,0.499){111}{\rule{0.711pt}{0.120pt}}
\multiput(828.00,386.17)(85.525,57.000){2}{\rule{0.355pt}{0.400pt}}
\multiput(915.00,444.58)(0.659,0.499){129}{\rule{0.627pt}{0.120pt}}
\multiput(915.00,443.17)(85.698,66.000){2}{\rule{0.314pt}{0.400pt}}
\multiput(1002.00,510.58)(0.588,0.499){145}{\rule{0.570pt}{0.120pt}}
\multiput(1002.00,509.17)(85.816,74.000){2}{\rule{0.285pt}{0.400pt}}
\multiput(1089.00,584.58)(0.518,0.499){163}{\rule{0.514pt}{0.120pt}}
\multiput(1089.00,583.17)(84.932,83.000){2}{\rule{0.257pt}{0.400pt}}
\multiput(1175.58,667.00)(0.499,0.523){171}{\rule{0.120pt}{0.518pt}}
\multiput(1174.17,667.00)(87.000,89.924){2}{\rule{0.400pt}{0.259pt}}
\multiput(1262.58,758.00)(0.499,0.569){171}{\rule{0.120pt}{0.555pt}}
\multiput(1261.17,758.00)(87.000,97.848){2}{\rule{0.400pt}{0.278pt}}
\sbox{\plotpoint}{\rule[-0.400pt]{0.800pt}{0.800pt}}%
\put(307,193){\usebox{\plotpoint}}
\put(307,190.84){\rule{20.958pt}{0.800pt}}
\multiput(307.00,191.34)(43.500,-1.000){2}{\rule{10.479pt}{0.800pt}}
\multiput(394.00,193.41)(3.881,0.511){17}{\rule{6.000pt}{0.123pt}}
\multiput(394.00,190.34)(74.547,12.000){2}{\rule{3.000pt}{0.800pt}}
\multiput(481.00,205.41)(1.915,0.505){39}{\rule{3.191pt}{0.122pt}}
\multiput(481.00,202.34)(79.376,23.000){2}{\rule{1.596pt}{0.800pt}}
\multiput(567.00,228.41)(1.378,0.503){57}{\rule{2.375pt}{0.121pt}}
\multiput(567.00,225.34)(82.071,32.000){2}{\rule{1.188pt}{0.800pt}}
\multiput(654.00,260.41)(1.096,0.502){73}{\rule{1.940pt}{0.121pt}}
\multiput(654.00,257.34)(82.973,40.000){2}{\rule{0.970pt}{0.800pt}}
\multiput(741.00,300.41)(0.930,0.502){87}{\rule{1.681pt}{0.121pt}}
\multiput(741.00,297.34)(83.511,47.000){2}{\rule{0.840pt}{0.800pt}}
\multiput(828.00,347.41)(0.808,0.502){101}{\rule{1.489pt}{0.121pt}}
\multiput(828.00,344.34)(83.910,54.000){2}{\rule{0.744pt}{0.800pt}}
\multiput(915.00,401.41)(0.703,0.502){117}{\rule{1.323pt}{0.121pt}}
\multiput(915.00,398.34)(84.255,62.000){2}{\rule{0.661pt}{0.800pt}}
\multiput(1002.00,463.41)(0.631,0.501){131}{\rule{1.209pt}{0.121pt}}
\multiput(1002.00,460.34)(84.491,69.000){2}{\rule{0.604pt}{0.800pt}}
\multiput(1089.00,532.41)(0.565,0.501){145}{\rule{1.105pt}{0.121pt}}
\multiput(1089.00,529.34)(83.706,76.000){2}{\rule{0.553pt}{0.800pt}}
\multiput(1175.00,608.41)(0.530,0.501){157}{\rule{1.049pt}{0.121pt}}
\multiput(1175.00,605.34)(84.823,82.000){2}{\rule{0.524pt}{0.800pt}}
\multiput(1263.41,689.00)(0.501,0.516){167}{\rule{0.121pt}{1.028pt}}
\multiput(1260.34,689.00)(87.000,87.867){2}{\rule{0.800pt}{0.514pt}}
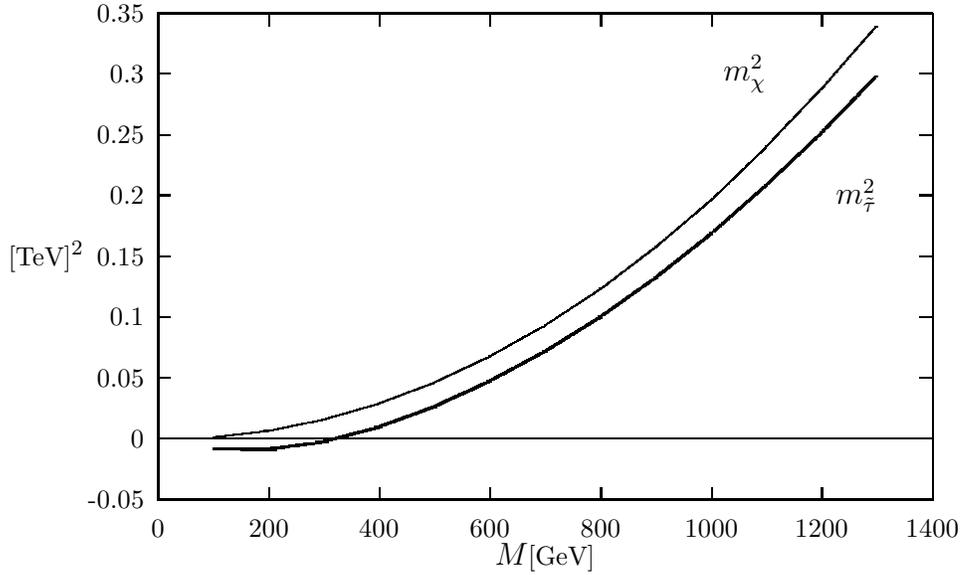
\end{picture} \\Fig. 9:   $m^2_{\tilde \tau}$
and $m^2_\chi$ as  function of $M$
for the universal choice.
\end{center}
As fig. 8 shows, $M$ has to be relatively large to satisfy
the constraint 
$m^2_{\tilde \tau} < m^2_\chi$ for the model ${\bf B}$. We
find, for this model too, that there is no region
of $M$ below $O$(few) TeV for the universal
choice in which
$m^2_{\tilde \tau} < m^2_\chi$ is satisfied.
In Table 4 we give a representative prediction
for the s-spectrum
for the model ${\bf B}$, where we have used:
$M=1$ TeV and $m_{{\bf 10}}= 0.65 $ TeV.
\begin{center}
Table 4:  A representative
example of the predictions of the s-spectrum
for the model ${\bf B}$.

\vspace{0.2cm}
\begin{tabular}{|c|c||c||c|}
 \hline
$m_{\chi}=m_{\chi_1}$ (TeV) &
0.44 &
$m_{\tilde{b}_2}$ (TeV)  &
1.79\\ \hline
$m_{\chi_2}$ (TeV) &
0.84 &
$m_{\tilde{\tau}}=m_{\tilde{\tau}_1}$ (TeV)  &
0.47
\\ \hline
$m_{\chi_3} $ (TeV) &
1.38 &
$m_{\tilde{\tau}_2}$ (TeV)   &
0.69
\\ \hline
$m_{\chi_4}$ (TeV) & 
1.39 &
$m_{\tilde{\nu}_1}$ (TeV)   &
0.62
\\ \hline
$m_{\chi^{\pm}_{1}} $ (TeV) &
0.84 &
$m_{A}$ (TeV) &
0.74
\\ \hline
$m_{\chi^{\pm}_{2}} $ (TeV) &
1.39 &
$ m_{H^{\pm}}$ 
(TeV & 
0.75
\\ \hline
$m_{\tilde{t}_1}$ (TeV) &
1.60 &
$ m_{H}$ 
(TeV &
0.74
\\ \hline
$m_{\tilde{t}_2}$  (TeV) &
1.82 &
$m_{h} $
(TeV) & 0.12
\\ \hline
$m_{\tilde{b}_1}$  (TeV) & 1.56
  & & 
\\  \hline
\end{tabular}

\end{center}

\vspace{0.2cm}

\section{Conclusion}
In this paper we have re-investigated the
two-loop finiteness conditions for the SSB parameters
in softly broken
$N=1$ supersymmetric Yang-Mills theories with a simple gauge group
and found that 
the previously known result \cite{jones2,jack3} on the $h=-M Y$ relation 
(\ref{hY}) is necessary while the universal solution
for the soft scalar masses can be continuously deformed
to the   sum rule (\ref{sumr}).

Since it has been known \cite{ibanez1,munoz1,jack1}  that
the universal soft scalar masses 
 appear for dilaton-dominated supersymmetry
 breaking in 4D superstring
models, we have  examined whether or not the 
two-loop corrected soft scalar-mass sum rule 
can also be obtained in some string model.  We have indeed found
that the same sum rule  is satisfied in 
a certain class of  string models  in which the massive string states are organized into $N=4$ supermultiplets so that
they do not contribute to the quantum modification
of the gauge kinetic function.
Since not only in finite GYU models,
but also in nonfinte GYU models  the same soft scalar-mass
sum rule is satisfied at least at the one-loop level \cite{kkk},
we believe
 there exists something non-trivial behind
these coincidences.

Motivated by these facts,
we have investigated the SSB sector  of two finite $SU(5)$ models
${\bf A}$ and ${\bf B}$.
We have found out that the two-loop corrections to the sum rule
is absent in these models. 
Since we do not know why this happens, it is an accident to us.
Finally we have investigated the low-energy sector of these models.
Using the sum rule and requiring that the LSP is  neutral, we have
constrained the  parameter space of the low-energy 
SSB sector in each model and
calculated the spectrum of the superparticles.
We have found that the model ${\bf A}$ allows relatively light
superparticles while in the model ${\bf B}$
they are heavier than $\sim 0.5$ TeV.
The mass of the lightest Higgs is $\sim 120$ GeV.

Taking into account all these results, we would like to
conclude that the finite models we have considered are 
not only academically attractive, but also 
making interesting predictions which are
consistent with
the present experimental knowledge.

\vspace{2cm}
\noindent
{\bf Acknowledgment}

\vspace{0.2cm}
We thank D. L\" ust and D. Suematsu for discussions.

\newpage

\begin{center}
{\bf Appendix}
\end{center}

\vspace{1cm}

The  RG functions
which we have used in the text are defined as:
$$
\frac{d}{dt}\,g = \beta_{g} =
\sum_{n=1}~\frac{1}{(16\pi^2)^n}\,\beta_{g}^{(n)}~,~ \frac{d}{dt}\,M =
\beta_{M} = \sum_{n=1}~\frac{1}{(16\pi^2)^n}\,\beta_{M}^{(n)}~,$$
$$\frac{d}{dt}\,Y^{ijk} = \beta_{Y}^{ijk}=Y^{ijp}\,
~\sum_{n=1}~\frac{1}{(16\pi^2)^n}\,\gamma_{p}^{(n)\,k} +(k
\leftrightarrow i) +(k\leftrightarrow j)~,$$
$$\frac{d}{dt}\,h^{ijk} =\beta_{h}^{ijk}=\sum_{n=1}~\frac{1}{(16\pi^2)^n}\,
[\beta_{h}^{(n)}]^{ijk}~,$$
$$\frac{d}{dt}\,(m^2)^{j}_{i} = [\beta_{m^2}]^{j}_{i}=
\sum_{n=1}~\frac{1}{(16\pi^2)^n}\,
[\beta_{m^2}^{(n)}]^{j}_{i}~,$$
$$\frac{d}{dt}\,b^{ij} = \beta_{b}^{ij}=
\sum_{n=1}~\frac{1}{(16\pi^2)^n}\,
\beta_{b}^{(n)ij}~,$$
where we assume that the gauge group is  a simple group.
The coefficients of the
 one-and two-loop RG functions  \cite{jones2,PW,rgf,jack1,jack3} are:
$$\beta_{g}^{(1)} = g^3\,[T(R)-3 C(G)]~,~
\beta_{M}^{(1)}=2 M\, \beta_{g}^{(1)}/g~, \eqno{(A.1)}$$
$$\gamma_{i}^{(1)\,j}=(1/2) Y_{ipq} Y^{jpq}
-2\delta^{j}_{i}\, g^2\,C(i)~, ~
\chi_{j}^{(1)i} = h^{imn}Y_{jmn}+4 M g^2 
C(i)\delta_{j}^{i}~,\eqno{(A.2)}$$
 $$
[\beta_{h}^{(1)}]^{ijk} = (1/2) h^{ijl} Y_{lmn} Y^{mnk}+
 Y^{ijl} Y_{lmn} h^{mnk}-2(h^{ijk}-2M Y^{ijk})\,g^2\,C(k) $$
$$+(k \leftrightarrow i)+(k \leftrightarrow j)~, \eqno{(A.3)}$$
$$
[\beta_{m^2}^{(1)}]^{j}_{i} = (1/2) Y_{ipq} Y^{pqn} (m^2)_{n}^{j}+
(1/2) Y^{jpq} Y_{pqn} (m^2)_{i}^{n}+
2Y_{ipq}Y^{jpr}(m^2)_{r}^{q}$$
$$+h_{ipq} h^{jpq} -8\delta_{i}^{j} M M^{\dag}\,g^2\,C(i)~,\eqno{(A.4)}$$
$$
\beta_{b}^{(1)ij}= b^{il} ~\gamma_{l}^{(1)\,j}
+\mu^{il}~ \chi_{l}^{(1)i}
+(i \leftrightarrow j),\eqno{(A.5)}$$
$$
\beta_{g}^{(2)}=
2 g^2  C(G)\beta_{g}^{(1)}-2 g^3 d^{-1}(G)\sum_i C(i) 
\gamma_{i}^{(1) i}~,\eqno{(A.6)}$$
$$\gamma_{j}^{(2)i} =
2 g  C(i)\delta_{j}^{i}\beta_{g}^{(1)}
-[~Y_{jmn}Y^{mpi}+2 g^2  C(j)
\delta_{j}^{p}\delta_{n}^{i}~] \gamma_{p}^{(1)n}~,\eqno{(A.7)}$$
$$\beta_{M}^{(2)}=
8 g  C(G)\beta_{g}^{(1)}M
+ g^2 d^{-1}(G)\sum_i C(i) ~[~-4\gamma_{i}^{(1) i}M+
 2\chi_{i}^{(1) i}~]~,\eqno{(A.8)}$$
$$\beta_{h}^{(2)ijk}=
-[~h^{ijl}Y_{lmn}Y^{mpk}+2Y^{ijl}Y_{lmn}h^{mpk}-4 g^2 M Y^{ijp} C(n)
\delta_{n}^{k}~] \gamma_{p}^{(1)n}$$
$$-2 g^2 [ h^{ijl} \gamma_{l}^{(1)k}+Y^{ijl} \chi_{l}^{(1)k}] C(k)
+g(2h^{ijk}-8 M Y^{ijk}) C(k) \beta_{g}^{(1)}$$
$$-Y^{ijl} Y_{lmn} Y^{pmk} \chi_{p}^{(1)n}+
(k \leftrightarrow i)+(k \leftrightarrow j)~,\eqno{(A.9)}$$
$$
\beta_{b}^{(2)ij}=
[~-b^{il}Y_{lmn}Y^{mpj}
-2\mu^{il} Y_{lmn}h^{pmj}-
Y^{ijl}Y_{lmn}b^{mp}
+4 g^2 M C(i)\mu^{ip}\delta_{j}^{n}
~]~ \gamma_{p}^{(1)n}$$
$$-[~\mu^{il}Y_{lmn}Y^{mpj}+\frac{1}{2} Y^{ijl}Y_{lmn}\mu^{mp} 
~] ~ \chi_{p}^{(1)n}
-2g^2 C(i) ~[~b^{il} \gamma_{l}^{(1)j}  +\mu^{il} \chi_{l}^{(1)j} ~]$$
$$+2  g C(i)\beta_{g}^{(1)}
~[~b^{ij}-2\mu^{ij}~]+(i \leftrightarrow j)~.\eqno{(A.10)}$$
$$
[\beta_{m^2}^{(2)}]^{j}_{i}=
-[~(m^2)^{l}_{i} Y_{lmn}Y^{mpj}
+\frac{1}{2}Y_{ilm}Y^{jpm}(m^2)^{l}_{n}+
\frac{1}{2}Y_{inm}Y^{jlm}(m^2)^{p}_{l}+
Y_{iln}Y^{jrp}(m^2)^{l}_{r}$$
$$ +h_{iln}h^{jlp} +4 g^2  |M|^2 C(j)
\delta_{n}^{j}\delta_{i}^{p}
+2 g^2 \sum_A (R_{A})^{j}_{i}(R_{A}m^2)^{p}_{n}
~]~ \gamma_{p}^{(1)n}$$
$$+[~ 2g^2  M^{\dag} C(i)
\delta_{n}^{j}\delta_{i}^{p}
- h_{iln} Y^{jlp} ~] ~ \chi_{p}^{(1)n}
-\frac{1}{2} [ Y_{iln} Y^{jlp}+2g^2 C(i)
\delta_{n}^{j}\delta_{i}^{p}] ~[\beta_{m^2}^{(1)}]^{n}_{p}$$
$$+4 |M|^2 C(i)\delta_{i}^{j}~ [ 3g \beta_{g}^{(1)}
+g^4 S']+\mbox{H.c.}~,\eqno{(A.11)}$$
where $S'$ is defined in eq. (\ref{S}).
Further references may be found for instance
in ref. \cite{barger}.

\end{document}